\documentclass[10pt,journal,compsoc]{IEEEtran}
\IEEEoverridecommandlockouts
\usepackage{cite}
\usepackage{tabularx}
\usepackage{amsmath,amssymb,amsfonts}
\usepackage{algorithmic}
\usepackage{graphicx}
\usepackage{textcomp}
\usepackage{xcolor}
\usepackage{listings}
\usepackage{balance}
\usepackage{amsmath}
\usepackage{amssymb}
\usepackage[justification=centering]{caption}
\usepackage{dcolumn}
\usepackage{array}
\usepackage{color}
\newcommand{\todo}[1]{}
\newcolumntype{d}[1]{D..{#1}}

\renewcommand{\todo}[1]{{\color{red} TODO: {#1}}}
\mathchardef \mhyphen="2D
\begin{document}

	\title{\textsc{DefectChecker}: Automated Smart Contract Defect Detection by Analyzing EVM Bytecode} 
	
	\author{Jiachi Chen, Xin Xia, David Lo, John Grundy, Xiapu Luo and Ting Chen
		\IEEEcompsocitemizethanks{\IEEEcompsocthanksitem Jiachi Chen, Xin Xia and John Grundy are with the Faculty of Information Technology, Monash University, Melbourne, Australia. \protect\\
			E-mail: \{Jiachi.Chen, Xin.Xia, John.Grundy\}@monash.edu
			\IEEEcompsocthanksitem David Lo is with the School of Information Systems, Singapore Management University, Singapore.\protect\\
			E-mail: davidlo@smu.edu.sg
			\IEEEcompsocthanksitem Xiapu Luo is with the Department of Computing, The Hong Kong Polytechnic University, Hong Kong.\protect\\
			E-mail: csxluo@comp.polyu.edu.hk
			\IEEEcompsocthanksitem Ting Chen is with the School of Computer Science and Engineering, University of Electronic Science and Technology of China, China.\protect\\
			E-mail: brokendragon@uestc.edu.cn
			\IEEEcompsocthanksitem Xin Xia is the corresponding author.}
		\thanks{Manuscript received     ; revised   }}

	\markboth{IEEE Transactions on Software Engineering, ~Vol.~  , No.~  , }%
	{Shell \MakeLowercase{\textit{et al.}}: Bare Demo of IEEEtran.cls for Computer Society Journals}

	\IEEEtitleabstractindextext{%
\begin{abstract}  
		Smart contracts are Turing-complete programs running on the blockchain. They are immutable and cannot be modified, even when bugs are detected. Therefore, ensuring smart contracts are bug-free and well-designed before deploying them to the blockchain is extremely important. A contract defect is an error, flaw or fault in a smart contract that causes it to produce an incorrect or unexpected result, or to behave in unintended ways. Detecting and removing contract defects can avoid potential bugs and make programs more robust. Our previous work defined 20 contract defects for smart contracts and divided them into five impact levels. According to our classification, contract defects with seriousness level between 1-3 can lead to unwanted behaviors, e.g., a contract being controlled by attackers. In this paper, we propose \textit{DefectChecker}, a symbolic execution-based approach and tool to detect eight contract defects that can cause unwanted behaviors of smart contracts on the Ethereum blockchain platform. \textit{DefectChecker} can detect contract defects from smart contracts' bytecode. We verify the performance of \textit{DefectChecker} by applying it to an open-source dataset. Our evaluation results show that \textit{DefectChecker} obtains a high F-score (88.8\% in the whole dataset) and only requires 0.15s to analyze one smart contract on average. We also applied \textit{DefectChecker} to 165,621 distinct smart contracts on the Ethereum platform. We found that 25,815 of these smart contracts contain at least one of the contract defects that belongs to impact level 1-3, including some real-world attacks.
	
\end{abstract}

		}
	\maketitle
	\IEEEdisplaynontitleabstractindextext
	
	
	\section{Introduction}
\label{Introduction}
In recent years, decentralized cryptocurrencies have attracted considerable interest. To ensure these systems are scalable  and secure without the governance of a centralized organization, decentralized cryptocurrencies adopt the \emph{blockchain} concept as their underlying technology.  Bitcoin\cite{bitcoin} was the first digital currency, and it allows users to encode scripts for processing transactions automatically. However, scripts in Bitcoin are not Turing-complete, which restricts their application to currencies, such as money transfer or payment. To address this limitation,  Ethereum~\cite{Ethereum_yellow_paper} leverages a technology named \emph{Smart Contracts}, which are Turing-complete programs that run on the blockchain. By utilizing this technology, practitioners can develop decentralized applications (DApps)~\cite{dapp} and apply blockchain techniques to different fields such as gaming~\cite{cryptokitties} and finance~\cite{ICO}.


 Smart contracts are usually developed using a high-level programming language, such as Solidity~\cite{Solidity}. When developers deploy a smart contract to Ethereum, the contract will first be compiled into Ethereum Virtual Machine (EVM) \textit{bytecode}. Then, each node on the Ethereum system will receive the smart contract bytecode and have a copy in their ledger.  Anyone, even attackers, can invoke the smart contract by sending transactions to the corresponding contract address. 

Key features of smart contracts make them become attractive targets for hackers~\cite{chen2017adaptive}. On the one hand, many smart contracts hold valuable Ethers, and they cannot hide their balance, which gives financial motivation for attacks by hackers~\cite{chen2018understanding, chen2020understanding}. On the other hand, smart contracts run in a permission-less network, which means hackers can check all the transactions and bytecode freely, and try to find bugs on the contracts. Even worse,  smart contracts cannot be modified, even when bugs are detected. Therefore, ensuring smart contracts are bug-free and well-designed before deploying them to Ethereum is extremely important. 

A contract defect~\cite{iso2017iec,fse-smell} is an error, flaw, or fault in a smart contract that causes it to produce an incorrect or unexpected result, or to behave in unintended ways~\cite{chillarege1996orthogonal}. The detection and removal of contract defects is a method to avoid potential bugs and improve the design of existing code. In our previous work~\cite{fse-smell}, we first defined 20 contract defects by analyzing StackExchange~\cite{StackExchange} posts.  It is also the first work that used an online survey to validate whether smart contract developers consider these contract defects as harmful, which make the definitions more persuasive. The work divided the defined 20 contract defects into five impact levels and showed that smart contracts contain defects with impact levels 1 to 3 can lead to unwanted behaviors, e.g., contracts being controlled by attackers. 

However, our previous work did not propose a suitable tool that could detect these contract defects. To address this limitation, in this paper, we propose \textit{DefectChecker} to detect eight contract defects defined in our previous work that belong to serious impact level 1 (high) to level 3 (medium), by using the bytecode of smart contracts. \textit{DefectChecker} symbolically executes the smart contract through bytecode, and without the needs of source code. During the symbolic execution, \textit{DefectChecker} generates the CFG of smart contracts, as well as the ``stack event", and identifies three features, i.e., ``Money Call", ``Loop Block", and ``Payable Function". By using the CFG, stack event, and the three features, we design eight rules to detect each contract defect.

We verify the performance of \textit{DefectChecker} by applying it to an open-source dataset developed in our previous work~\cite{fse-smell}. We also compare its results with those of three state-of-the-art tools, i.e., \textit{Oyente, Mythril} and \textit{Securify}. Our evaluation results show that \textit{DefectChecker} obtains the highest F-score (88.8\% in the whole dataset) and requires the least time (0.15s per contract) to analyze one smart contract compared to these other baseline tools. We also crawled all of the bytecode of smart contracts deployed on Ethereum by Jan. 2019 and applied \textit{DefectChecker} to these \textit{165,621} distinct bytecode smart contracts. We found that \textit{15.9\%} of smart contracts on Ethereum contain at least one of contract defects (the severity level 1 to 3 ) using \textit{DefectChecker}. 


The main contributions of this work are: 

\begin{itemize}
	\item To the best of our knowledge, \textit{DefectChecker} is the most \textbf{accurate} and the \textbf{fastest} symbolic execution-based model for smart contract defects detection. 
	
	\item We systematically evaluated our tool using an open source dataset to test its performance. In addition, we  crawled all of the bytecode (165,621) on the Ethereum platform by the time of writing the paper and identified 25,815 smart contracts that contain at least one contract defect. Using these results, we find some real-world attacks, and give examples to show the importance of detecting contract defects.
	
	\item Our datasets, tool and analysis results have been released to the community at https://github.com/Jiachi-Chen/DefectChecker/.  
	
\end{itemize}

The organization of the rest of this paper is as follows. In Section 2, we provide background knowledge of smart contracts and introduce eight contract defects with code examples. Then, we introduce the architecture of \textsc{DefectChecker} in section 3 and present its evaluation in section 4. We conduct a large scale evaluation based on Ethereum smart contracts in Section 5 and give two real-world attacks as case studies. In section 6, we introduce the related works. Finally, we conclude the study and discuss possible future work in Section 7. 

	\section{Background and Motivation}
\label{background}
In this section, we briefly introduce key background information about smart contracts and their contract defects.

\subsection{Smart Contracts}
\noindent {\bf Contracts.}
Leveraging blockchain techniques, smart contracts are autonomous protocols stored on the blockchain. Once started, the running of a contract is automatic and it runs according to the program logic defined beforehand~\cite{chen2018towards}.  When developers deploy a smart contract to Ethereum, the contract will be compiled to EVM bytecode and identified by a unique 160-bit hexadecimal hash contract address. The smart contract execution depends on their code, and even the creator cannot affect its running or state. For example, if a contract does not contain functions for Ether transfer, even the creator cannot withdraw the Ethers. Smart contracts run on a permission-less network. Anyone can invoke the methods of smart contracts through ABI (Application Binary Interface)~\cite{Solidity}. The contract bytecode, transactions, and invocation parameters are visible to everyone.

\noindent {\bf Gas System.}
To ensure the security of smart contracts, each transaction of a smart contract will be run by all miners.  Ethereum uses the \textit{gas system}~\cite{whitepaper} to measure its computational effort, and the developers who send transactions to invoke smart contracts need to pay an execution fee. The execution fee is computed by: $gas\_cost \times gas\_price$. Gas cost depends on the computational resource that takes by the execution and gas price is offered by the transaction creators. To limit gas cost, when developers send their transactions to invoke contracts, they will set the \textit{Gas Limit} which determines the maximum gas cost. If the gas cost of a transaction exceeds its \textit{Gas Limit}, the execution will fail and throw an \emph{out-of-gas error}~\cite{Ethereum_yellow_paper}. There are some special operations which will limit the \textit{Gas Limit} to a specific value. For example, \emph{address.transfer()} and \emph{address.send()} are two methods provided by Ethereum that are used to send Ethers. If a smart contract uses these methods to send Ethers to another smart contract, the \textit{Gas Limit} will be restricted to 2300 gas units~\cite{Solidity}. 2300 gas units are not enough to write to storage, call functions or send Ethers, which can lead to the failure of transactions. Therefore, \emph{address.transfer()} and \emph{address.send()} can only be used to send Ethers to \textit{external owned accounts} (EOA). (There are two types of accounts on Ethereum: externally owned accounts which controlled by private keys, and contract accounts which controlled by their contract code~\cite{Ethereum_yellow_paper}.)

\noindent {\bf Ethereum Virtual Machine (EVM).}
To deploy a smart contract to Ethereum, its source code needs to be compiled to bytecode and stored on the blockchain. EVM is a stack-based machine; when a transaction needs to be executed, EVM will first split bytecode into bytes; each byte represents a unique instruction called opcode. There are 140 unique opcodes by April 2019~\cite{Ethereum_yellow_paper}, and each opcode is represented by a hexadecimal number~\cite{Ethereum_yellow_paper}. EVM uses these opcodes to execute the task. For example, consider a bytecode 0x6070604001. EVM first splits this bytecode into bytes (0x60, 0x70, 0x60, 0x40, 0x01), and execute the first byte 0x60, which refers to opcode \textit{PUSH1}. \textit{PUSH1} pushes one byte data to EVM stack. Therefore, 0x70 is pushed to the stack. Then, EVM reads the next 0x60 and push 0x40 into the stack. Finally, EVM executes 0x01, which refers to opcode \textit{ADD}. \textit{ADD} obtains the next two values from the top of the stack,  i.e., 0x70 and 0x40, and put their sum (B0), a hex result into the stack.

\noindent {\bf EVM Bytecode v.s. JVM Bytecode in Control Flow Analysis.}
Control flow analysis methods have been widely used in other stack-based machines, e.g., JVM~\cite{soot}. However, there are many differences in analyzing the control flow of Java bytecode and EVM bytecode. These differences present some new challenges in analyzing EVM bytecode. We highlight the key differences between EVM bytecode analysis method we used in this paper and JVM bytecode analysis. These include:

 (1) JVM bytecode has a fixed stack depth under different control-flow paths. The execution of JVM cannot reach the same program point with different stack sizes~\cite{grech2019gigahorse}. There are no such constraints for EVM bytecode, which greatly increases the difficulty of identifying the control-flow constructs in EVM bytecode. For example, for a simple recursive code ``function f(int a){f(a);}". The code will be compiled in EVM as:

 \begin{lstlisting}
 Block 1: 
	 JUMPDEST            
	 PUSH Block1' ID    
	 DUP2                     
	 PUSH Block2' ID     
	 JUMP                    
 Block 2:
 	JUMPDEST         
 \end{lstlisting}
 
 There are two blocks; two block identifiers are pushed in the same block (block 1) and will be read by the same instruction (JUMP). The difference between the JVM and EVM is that the JVM creates a new frame~\cite{jvm} with a new operand stack for each method call, whereas the EVM just has one global operand stack. (A frame is used to store data and partial results, as well as to perform dynamic linking, return values for methods, and dispatch exceptions.)

 
  (2) JVM bytecode has a clearly defined set of targets for each jump~\cite{chen2019large}. In contrast, the jump target for EVM bytecode is read from the EVM stack. When a conditional jump is used, the target will be affected by the second stack item. For example, in Figure 2, the jump target of JUMPI (ID 140) is read from previous instruction PUSH and will be affected by the second stack item, i.e., \textit{ISZERO(GT(10, num))} (details see Section 3.3). If the second item refers to a true value. The jump target is 148; otherwise, the target is 141. The unconditional jump target is also read from the top of the EVM stack. For example, the jump target of \textit{JUMP} (ID 147) in Figure 2 is also read from the previous instruction \textit{PUSH}. Therefore, we need to symbolically execute the EVM bytecode to construct the control-flow edges.

 (3) JVM bytecode has well-defined method invocation and return instructions~\cite{grech2019gigahorse}. In contrast, EVM bytecode uses jumps to perform its intra-contract function calls. In this case, to resolve an intra-contract function call, we need to inspect the top stack element to determine the jump target. For example, there are two functions A and B. Function A contains three blocks, e.g., A1, A2, A3; function B contains two blocks, e.g., B1, B2. The code on block A2 calls function B. In EVM bytecode, there is no defined method invocation and return instructions. Instead, the code pushes the return address to the stack; the arguments and jump target (block identifier of B1) need to be identified through bytecode. To return, the code pops the caller's block identifier (A3) and jumps to execute the block. Thus, the execution sequences are A1, A2, B1, B2, A3. The identifiers of B1 and A3 should be obtained from bytecode through symbolic execution.     



\noindent {\bf The Fallback Function.}
The fallback function is a unique feature of smart contracts compared to traditional programs. An example can be found at Line 13 of Listing 4, which is the only unnamed function in smart contracts programming~\cite{Solidity}. The fallback function does not have any arguments or return values. It will be executed automatically on a call to the contract if none of the functions match the given function identifier~\cite{Solidity}. For example, if a transaction calls  function `A' of the contract, and there is no function named `A', then the fallback function will automatically be executed to handle the erroneous function invocation. If the function is marked by \textit{payable}~\cite{Solidity}, the fallback function will also be executed automatically when receiving Ethers. 

\noindent {\bf The Call Instruction and Ether Transfer.}
Ether transfer is an important feature on Ethereum. In Solidity programming, there are three methods to transfer Ethers, i.e., \textit{address.call.value()}, \textit{address.transfer()}, and \textit{address.send()}. Among these three methods, only \textit{address.call.value()} allows users to send Ethers to a contract address, as the other two methods are limited to 2300 gas units, which are not enough to send Ethers. \textit{address.send()} returns a boolean value, while \textit{address.transfer()} throws an exception when errors happen and returns nothing. All of these three methods can generate a CALL instruction in contract bytecode. Other behaviors, e.g., function call, can also generate CALL instructions. A CALL instruction reads seven values from the top of EVM stack. They represent the gas limitation, recipient address, transfer amount, input data start position, size of the input data,  output data start position, size of the output data, respectively.

\subsection{Contract Defects in Smart Contracts}
Our previous work \cite{fse-smell} defined 20 contract defects for smart contracts. We divided these contract defects into five ``impact" levels; among these contract defects, 11 belong to impact level one (most serious) to three (low seriousness) that might lead to unwanted behaviors. The definition of these 11 contract defects is given in Table~\ref{tab:difinition}. In this paper, we propose \textit{DefectChecker}, a symbolic execution tool to detect eight of these impact level one to three contract defects. \textit{DefectChecker} does not detect contract defects belonging to levels 4 and 5, as these contract defects will not affect the normal running of the smart contracts according to the definition. For example, \textit{Unspecified Compiler Version} is one of the level 5 smart contract defects. The removal of the contract defects requires the developer of the contract to use a specific compiler like 0.4.25. This contract defect will not affect the normal running of the contract and will only pose a threat for code reuse in the future. This kind of contract defect is also difficult to  detect at the bytecode level as much semantic information is lost after compilation. 

However, please note that in this work, we do not consider three of the contract defects that belong to impact level 1 to 3 -- \textit{Unmatched Type Assignment}, \textit{Hard Code Address} and \textit{Misleading Data Location}, as they are not easy to detect at  bytecode level. Our analysis shows that they appear 22, 84, and 1 times among 587 smart contacts, respectively. EVM will remove or add some information when compiling smart contracts to bytecode, which may cover up these taints on the source contract code. For \textit{\textbf{Hard Code Address}}, the bytecode we obtain from the blockchain does not contain information on the \textit{construct} function, while we found most \textit{Hard Code Address} errors appear in \textit{construct} functions. To detect \textit{\textbf{Unmatched Type Assignment}}, we need to know the maximum loop iterations, which is usually read from storage, and is not easy to obtain the value through static analysis. For example, for a loop ``\textit{for(uint8 i = 0; i $\textless$ num; i++)}", the data range of uint8 is from 0 to 255. Thus, if num is larger than 255, the loop will overflow. However, num is usually a storage variable which is read from storage or depends on an external input. Thus, it is difficult to detect this through bytecode analysis. \textit{\textbf{Misleading Data Location}} is also not easy to detect from bytecode. In Solidity programming, \textit{storage} in Solidity is not dynamically allocated and the type of \textit{struct, array} or \textit{mapping} are maintained on the storage. Thus, these three types created inside a function can point to the \textit{storage slot} 0 by default, which can lead to potential bugs. However, we cannot know whether the point on slot 0 is correct or a mistake made by EVM.

\begin{table*} 
	\scriptsize
	\caption{The Definitions of contract defects with Impact level 1-3. The first eight contract defects can be detected by \textit{DefectChecker}.}
	\label{tab:difinition}
	\begin{center}
		\begin{tabular}{p{65pt} | p{140pt} |p{20pt} <{\centering} || p{65pt} | p{140pt} |p{20pt}<{\centering}}
			\hline
			\textbf{Contract Defect} & \textbf{Definition} & \textbf{Impact Level} & \textbf{Contract Defect} & \textbf{Definition} & \textbf{Impact Level}\\
			\hline
			\textit{Transaction State Dependency (TSD)}  & Using tx.origin to check the permission. & IP1  & \textit{DoS Under External Influence (DuEI)}  & Throwing exceptions inside a loop which can be influenced by external users& IP2 \\
			\hline
			\textit{Strict Balance Equality (SBE)}   & Using strict balance quality to determine the execute logic. & IP2 &  \textit{Reentrancy (RE)} & The re-entrancy bugs. & IP1 \\
			\hline
			\textit{Nested Call (NC)}  & Executing CALL instruction inside an unlimited-length loop.& IP2  &  \textit{Greedy Contract (GC)} & A contract can receive Ethers but can not withdraw Ethers. & IP3 \\
			\hline
			\textit{Unchecked External Calls (UEC)} & Do not check the return value of external call functions. & IP3 &   \textit{Block Info Dependency (BID)} & Using block information related functions to determine the execute logic. & IP3 \\
			\hline
			\hline
			\textit{Unmatched Type Assignment} & Assigning unmatched type to a value, which can lead to integer overflow & IP2 & \textit{Misleading Data Location} & The reference types of local variables with \textit{struct}, \textit{array} or \textit{mapping} do not clarify & IP2 \\
			\hline
			\textit{Hard Code Address} & Using hard code address inside smart contracts. & IP3 \\
			\hline
		\end{tabular}
	\end{center}
\end{table*}

\subsubsection{Definition of Impact Levels}

Below we give representative concrete examples of each of the eight smart contract defects, and introduce the definition of impact level one to three according to our previous work.

\begin{itemize}
\item \textbf{Impact 1 (IP1):}  Smart contracts containing these contract defects can lead to critical unwanted behaviors. Unwanted behaviors can be triggered by attackers, and they can make profits by utilizing the defects.
\item \textbf{Impact 2 (IP2):} Smart contracts containing these contract defects can lead to critical unwanted behaviors. Unwanted behaviors can be triggered by attackers, but they cannot make profits by utilizing the defects.
\item \textbf{Impact 3 (IP3):} There are two types of IP3. \textbf{Type A}: Smart contracts containing these contract defects can lead to critical unwanted behaviors, but unwanted behaviors cannot be triggered by attackers. \textbf{Type B}: Smart contracts containing these contract defects can lead to major unwanted behaviors. The unwanted behaviors can be triggered by attackers, but they cannot make profits by utilizing the defects.
\end{itemize}	


Critical represents contract defects, which can lead to a crash, being controlled by attackers, or can lose all the Ethers. Major represents the contract defects that can lead to the loss of a part of the Ethers ~\cite{fse-smell}.

\subsubsection{Examples of Smart Contract Defects}

\begin{lstlisting}[caption={Transaction State Dependency}]
contract Victim { ...
  address owner = owner_address;
  function sendMoney(address addr){
	require(tx.origin == owner);
	addr.transfer(1 Ether); 
  } 
}
contract Attacker{ ...
  function attack(address vim_addr,address myAddr){
	Victim vic = Victim(vim_addr);
	vic.sendMoney(myAddr);
 }
}
\end{lstlisting}

\textbf{\textit{(1). Transaction State Dependency (TSD):}} Contracts need to check whether the caller has the right permission for some permission sensitive functions. The failure of the permission check can cause serious consequences. \textit{tx.origin} can get the original address of the transaction, but this method is not reliable as the address returned by this method depends on the transaction state. Therefore, \textit{tx.origin} should not be used to check whether the caller has permission to execute functions.

\textbf{Example}:  In Listing 1, The \textit{Attacker} contract can make a permission check fail by utilizing the \textit{attack} function (Line 9). By utilizing this method, anyone can execute \textit{sendMoney} function (Line 3) and withdraw the Ethers in the contract.

\textbf{Possible Solution}: \textit{Solidity} provides \textit{msg.sender} to obtain the sender address, which can be used to check permissions instead of using \textit{tx.origin}.

\textbf{\textit{(2). DoS under External Influence (DuEI):}} Smart contracts will rollback a transaction if exceptions are detected during their running.  If the error that leads to the exception cannot be fixed, the function will give a denial of service (\textit{DoS}) error perpetually.

\textbf{Example}: Listing 2 shows such an example. Here, \textit{members} is an array which stores many addresses. However, one of the address is an attacker contract, and the transfer function can trigger an out-of-gas exception due to the 2300 gas limitation~\cite{Ethereum_yellow_paper}. Then, the contract state will rollback. Since the code cannot be modified, the contract can not remove the attack address from \textit{members} list, which means that if the attacker does not stop attacking, the following function cannot work anymore.

\textbf{Possible Solution}:  Developers can use a boolean value check instead of throwing exceptions in the loop. For example, using ``\textit{if(members[i].send(0.1 ether) == false) break;}" instead of line 3 in listing 2.

\begin{lstlisting}[caption={DoS under External Influence}]
for(uint i = 0;i < members.length; i++){
	if(this.balance > 0.1 ether)
		members[i].transfer(0.1 ether); 
}
\end{lstlisting}

\textbf{\textit{(3). Strict Balance Equality (SBE): }}Attackers can send Ethers to any contracts forcibly by utilizing \textit{selfdestruct()}~\cite{Solidity}. This method will not trigger the fallback function, which means the victim contract cannot reject the Ethers. Therefore, smart contract logic may fail to work due to the unexpected Ethers sent by attackers.  

\textbf{Example}:  The \textit{doingSomething()} function in listing 3 can only be triggered when the balance strict equal to 1 \textit{ETH}. However, the attacker can send 1 \textit{Wei} (1 ETH = 1e18 Wei) to the contract to make the balance never equal to 1 ETH.

\textbf{Possible Solution}: The contract can use ``$\geq$" to replace ``==" as attackers can only add to the amount of a balance. In this case, it is difficult for the attackers to affect the logic of the program. 

\begin{lstlisting}[caption={Strict Balance Equality:}]
if(this.balance == 1 eth)  doingSomething();
\end{lstlisting}

\textbf{\textit{(4). Reentrancy (RE):}} In Ethereum, a function can be executed several times in one execution by using the \textit{Call} method. When a contract calls another, the execution waits for the call to finish~\cite{oyente}. Thus, it can lead to multiple invocations and money transfer in some situations. 

\textbf{Example}: Listing 4 shows an example of a reentrancy defect. There are two smart contracts, i.e., \textit{Victim} contract and \textit{Attacker} contract. The \textit{Attacker} contract is used to transfer Ethers from \textit{Victim} contract, and the \textit{Victim} contract can be regarded as a bank, which stores the Ethers of users. Users can withdraw their Ethers by invoking \textit{withdraw()} function, which contains Reentrancy defects. 

First, the \textit{Attacker} contract uses the \textit{reentrancy()} function (L16) to invoke \textit{Victim} contract’s \textit{withdraw()} function in line 3. The \textit{addr} in line 16 is the address of the \textit{Victim} contract. Normally, the \textit{Victim} contract sends Ethers to the callee in line 6, and resets callee’s balance to 0 in line 7. However, the \textit{Victim} contract sends Ethers to the \textit{Attacker} contract before resetting the balance to 0. When the \textit{Victim} contract sends Ethers to the \textit{Attacker} contract (L6), the fallback function (L13) of the \textit{Attacker} contract will be invoked automatically, and then invoking the \textit{withdraw()} function (L14) again.  The invoking sequence in this example is: L16-17 $\rightarrow$ L3-6 $\rightarrow$ L13-14 $\rightarrow$  L3-6 $\rightarrow$ L13-14 $\cdots$, until Ethers run out.

\begin{lstlisting}[caption={Reentrancy}]
contract Victim { 	...
  mapping(address => uint) public userBalance;
  function withdraw(){
	uint amount = userBalance[msg.sender];
	if(amount > 0){
	  msg.sender.call.value(amount)();
	  userBalance[msg.sender] = 0; 
	  } 
  } 
  ... 
}
contract Attacker{ 	...
  function() payable{
	Victim(msg.sender).withdraw(); 
  }
  function reentrancy(address addr){
	Victim(addr).withdraw();
  } 
  ...
}
\end{lstlisting}

\textbf{Possible Solution}: There are 3 kinds of \textit{Call} methods that can be used to send Ethers in Ethereum, i.e., \textit{address.send()}, \textit{address.transfer()}, and \textit{address.call.value()}. \textit{address.send()} and  \textit{address.transfer()} will change the maximum gas limitation to 2300 gas units if the recipient is a contract account. 2300 gas units are not enough to transfer Ethers, which means \textit{address.send()} and  \textit{address.transfer()} cannot lead to \textit{Reentrancy}. Therefore, using \textit{address.send()} and  \textit{address.transfer()}  instead \textit{address.call.value()} can avoid \textit{Reentrancy}.

\textbf{\textit{(5). Nested Call (NC):}} Instruction \textit{CALL} is very expensive (9000 gas paid for a non-zero value transfer as part of the CALL operation)~\cite{Ethereum_yellow_paper}. If a loop contains the \textit{CALL} instruction but does not limit the loop iterations,  the total gas cost may have a high risk to exceed its gas limitation.  

\textbf{Example}: In listing 5, if we do not limit the loop iterations, attackers can maliciously increase its size to cause an out-of-gas error.  Once the out-of-gas error happens, this function cannot work anymore, as there is no way to reduce the loop iterations.

\textbf{Possible Solution}: Developers should estimate the maximum loop iterations and limit the loop iterations.

\begin{lstlisting}[caption={Nested Call}]
for(uint i = 0; i < member.length; i++){
	member[i].send(1 wei); 
}
\end{lstlisting}

\textbf{\textit{(6). Greedy Contract (GC):}} Ethers on smart contracts can only be withdrawn by sending Ethers to other accounts or using \textit{selfdestruct} function. Otherwise, even the creators of the smart contracts cannot withdraw the Ethers and Ethers will be locked forever. We define that a contract is a greedy contract if the contract can receive Ethers (contains payable functions) but there is no way to withdraw the Ethers. 

\textbf{Example}: Listing 6 is a greedy contract. The contract is able to receive Ethers as it contains a payable fallback function in line 2. However, the contract does not contain any methods to transfer money to others. Therefore, the Ethers on the contract will be locked forever. 

\textbf{Possible Solution}:  Adding a function to withdraw Ethers if the contract can receive Ethers.

\begin{lstlisting}[caption={Greedy Contract}]
Contract Greedy{
	function() payable{ 
		process(msg.sender); 
	}
	function process(address addr) {...}
}
\end{lstlisting}

\textbf{\textit{(7). Unchecked External Call (UEC): }}  Solidity provides many functions (\textit{address.send(), address.call()}) to transfer Ethers or call functions between contracts. However, these call-related methods can fail, e.g., have a network error or run out of gas. When errors happen, these functions will return a boolean value but never throw an exception. If the callers do not check the return values of the external calls, they cannot ensure whether the logic of the following code snippets is correct. 

\textbf{Example}: Listing 7 shows such an example. Line 1 does not check the return value of the \textit{address.send()}. As the Ether transfer can sometimes fail, line 1 cannot ensure whether the logic of the following code is correct. 

\textbf{Possible Solution}:  Always checking the return value of the \textit{address.send()} and \textit{address.call()}.
\begin{lstlisting}[caption={Unchecked External Call}]
address.send(ethers); doingSomething();  //bad
if(address.send(ethers)) doingSomething(); //good
\end{lstlisting}

\textbf{\textit{(8). Block Info Dependency (BID):}} Developers can utilize a series of block related functions to obtain block information. For example, block.blockhash is used to obtain the hash number of the current block.  Many smart contracts rely on these functions to decide a program's execution, e.g., generating random numbers.  However, miners can influence block information, e.g, miners can vary the block time stamp by roughly 900 seconds~\cite{oyente}. In this case, the block info dependency operation can be controlled by miners to some extent.

\textbf{Example}: The contract in listing 8 is a code snippet of a roulette contract. The contract utilizes block hash number to select a winner, and send winner one Ether as bonus. However, the miner can control the result. So, the miner can always be the winner. 

\textbf{Possible Solution}: The precondition of a safe random number is that the random number cannot be controlled by a single person, e.g., a miner.  The completely random information we can use in Ethereum includes users' addresses, users' input numbers and so on. Also, it is important to hide the values used by the contract for other players to avoid attacks. Since we cannot hide the address of users and their submitted values on Ethereum, a possible solution to generate random numbers without using block related functions is using a hash number. The algorithm has three rounds:

\textit{Round 1}: Users obtain a random number and generate a hash value in their local machine. The hash value can be obtained by \textit{keccak256} function, which is a function provided by Ethereum. After obtaining the random number, users submit the hash number.

\textit{Round 2}: After all the users submit the hash number, users are required to submit the original random number. The contract checks whether the original number can generate the same hash number by using the same \textit{keccak256} function.

\textit{Round 3}: If all users submit correct original numbers, the contract can use the original numbers to generate a random number.

\begin{lstlisting}[caption={Block Info Dependency: }]
address[] participators;
uint winnerID = uint(block.blockhash) % participators.length
participators[winnerID].transfer(1 eths);
\end{lstlisting}

	\section{The \textit{DefectChecker} Approach} 

\subsection{Design Overview} 
\begin{figure}
	\begin{center}
		\includegraphics[width=0.45\textwidth]{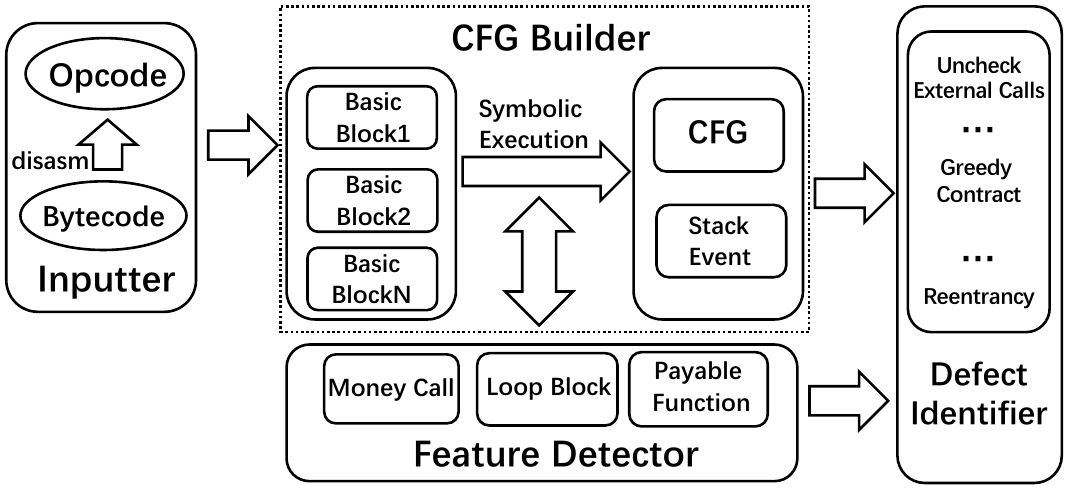} \vspace{-0.1cm}
		\caption {Overview architecture of \textit{DefectChecker}} 
		\label{Fig:overall}
	\end{center}
\end{figure}

Figure~\ref{Fig:overall} depicts an overview architecture of the \textit{DefectChecker} approach. There are four components of \textit{DefectChecker}, i.e., \textit{Inputter}, \textit{CFG Builder}, \textit{Feature Detector}, and \textit{Defect Identifier}. 

The left part of the figure is the Inputter, and users can feed bytecode as \textit{input}. Solidity source code is also allowed, but it needs to be compiled into bytecode.  Bytecode is then disassembled into opcodes by utilizing API provides by \textit{Geth}~\cite{geth}. Then, \textit{DefectChecker} splits opcode into several basic blocks and symbolically executes instructions in each block. After that, \textit{DefectChecker} generates the CFG (control flow graph) of a smart contract and records all stack events. During symbolic execution, \textit{Feature Detector} detects three features (i.e., Money Call, Loop Block and Payable Function), all concepts introduced below.  Based on this information, \textit{Defect Identifier} uses eight different rules to identify the contract defects on smart contracts.

Detecting contract defects by bytecode is very important for smart contracts on Ethereum. All the bytecode of smart contracts are stored on the blockchain, but only less than 1\% of smart contracts have opened their source code~\cite {TokenScope}. Smart contracts usually call other contracts, but the callee contracts may not open their source code for inspection. In such a case, the caller smart contracts can only detect whether the callee contract is secure through their bytecode.

\vspace{-0.3cm}

\subsection{Basic Block Builder} 

A basic block is a straight-line code sequence with no branches in except to the entry and no branches out except at the exit~\cite{chen2017under}. We first split the opcode into several blocks and give a type of the block according to its exit type. The exit type can be determined by the last instruction on a block. If the last instruction is \textit{JUMP} or \textit{JUMPI}, the block type is \textit{unconditional} or \textit{conditional}, respectively. If the last instruction is a terminal instruction (\textit{STOP}, \textit{REVERT} and \textit{RETURN}), the block type is \textit{terminal}. Some blocks belong to none of these three types, we call their block type  as \textit{fall}. In summary, we consider four types of blocks: \textit{unconditional}, \textit{conditional}, \textit{fall}, and \textit{terminal}.

\subsection{Symbolic Execution}
\label{lab:se}
Unlike other stack-based machines, e.g., JVM where  Java bytecode has a clearly-defined set of targets for every jump, the jump position of EVM bytecode needs to be calculated during symbolic execution. Thus, \textit{DefectChecker} needs to symbolically execute each single EVM instruction one at a time to obtain the CFG for smart contracts.  EVM is a stack-based machine -- when executing an instruction, it reads several symbolic states from the top of the EVM stack and put the symbolic result back to the EVM stack. During the symbolic execution, we can obtain the jump relations between blocks. There are three types of block according to the jump behaviors, i.e., \textit{conditional jump}, \textit{unconditional jump} and \textit{fall execution}. \textit{Stack Event} records all symbolic states on the EVM stack after the execution of each instruction.

\begin{lstlisting}[caption={Code of Figure 2}]
function example(uint num) returns(uint){
	if(num > 10)
		return 1;
	else{
		return 0;
	}
} 
\end{lstlisting}

\begin{figure}
	\begin{center}
		
		\includegraphics[width=0.49\textwidth]{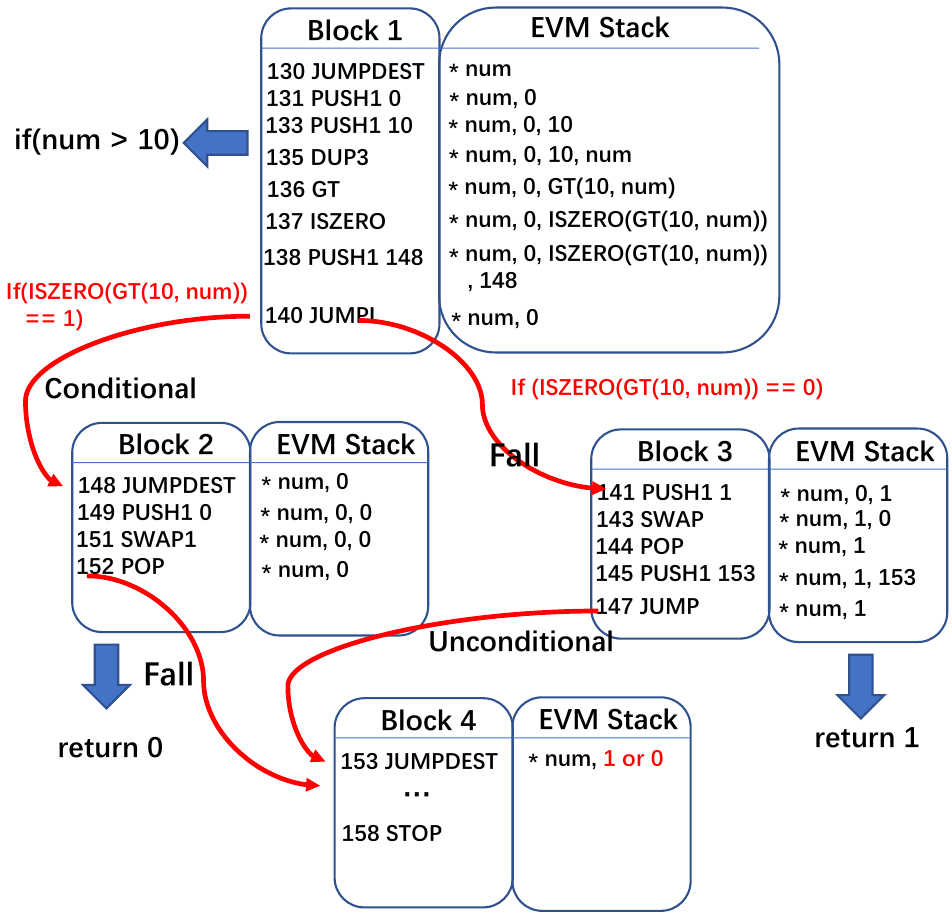} 
		\caption {Example of Symbolic Execution} 
		\label{Fig:se}
	\end{center} 
\end{figure}

Figure~\ref{Fig:se} is an example of the symbolic execution of the code in Listing 9. There are 4 blocks in this figure, and each block contains several instructions. The instructions in block 1 represent the code \textit{if(num \textgreater 10)}. The block 2 and block 3 put the value (0 or 1) to the EVM stack, respectively. The instructions in \textit{block 4}  are used to return the value(0 or 1) to the environment. The left-most number in each line indicates instructions' index ID, and the center part is the instruction that needs to be executed. All the instructions will execute sequentially according to their index ID. If the instruction is `PUSH', the right-most part will have a value that pushes into EVM stack. There is a Program Counter (PC) that records the ID that being executed at the current time. The PC starts from ID 0 in block 1, and EVM executes this instruction. 

The example shown in Figure~\ref{Fig:se} is a part of the code of a contract, so the PC starts from index ID 130 in block 1. Before EVM executes the instruction \textit{JUMPDEST}, there is a symbol \textit{num} in the EVM stack. The symbol \textit{num} represents the input value of the function (L1 of Listing 9).  \textit{JUMPDEST} marks a valid destination for jumps; it does not read or push any values. So the PC points to ID 131, and EVM pushes a value 0 to EVM stack. Then, `10' is pushed into EVM stack and PC point to 135. \textit{DUP3} duplicates the 3rd stack item. Therefore, the symbol \textit{num} is pushed into EVM stack. \textit{GT} reads two values from the EVM stack. If the first value (at the top of the stack) is greater than the second value, than EVM push 1 into the stack; otherwise, 0 is pushed. We use a symbol \textit{GT(a, num)} to represent the result and push the result into the EVM stack. Then,  \textit{ISZERO} reads a value from the top of the EVM stack. ISZERO reads one value from EVM. If the value equal to zero, then we push 1 into stack; otherwise, we push 0. We use a symbol \textit{ISZERO(GT(a, num))} to represent the result and push the result into the EVM stack. \textit{JUMPI} (ID 140) reads two values from the stack, the first value represents the jump position `148', and the second value is a conditional expression. If the result of the conditional expression is ``1" (true), the the PC jumps to the index ID 148, which indicates the start position of block 2. Otherwise, if the result is ``0" (false), the EVM falls to execute the following index ID 141(the start position of Block 3). 

Since the result of \textit{ISZERO(GT(a, num))} can be ``0" or ``1", this symbolic execution can generate two paths, i.e., Block 1 $\rightarrow$ Block 2 and Block 1$\rightarrow$ Block 3. 

We first assume the result of \textit{ISZERO(GT(a, num))} is ``1" and the path is Block 1 -\textgreater Bock 2. In this case, the PC points to the ID 148. The jump type of this path is \textit{conditional jump}. After executing the instructions on ID 148-152, the EVM falls to execute block 4. The jump type from block 2 to block 4 is \textit{fall}. When executing the first instruction of the block 4, the EVM stack holds two values, i.e., \textit{num} and \textit{0}. Block 4 then returns the value 0 to the environment and uses instruction \textit{STOP} to finish the execution. 

We then assume the result of \textit{ISZERO(GT(a, num))} is ``0" and the path is Block 1 -\textgreater Bock 3. In this case, the PC points to the ID 141. The jump type of this path is \textit{fall execution}. \textit{JUMP} refers to an unconditional jump; it reads one value from the top of the stack. The value reads by \textit{JUMP} in ID 147 is `153'. After executing the instructions on ID 141-147, the EVM then jumps to execute block 4. The jump type from block 3 to block 4 is an \textit{unconditional jump}. When executing the first instruction of the block 4, the EVM stack holds two values, i.e., \textit{num} and \textit{1}. Block 4 then returns the value 1 to the environment and uses instruction \textit{STOP} to finish the execution.


When executing a conditional jump, we should determine the satisfiability of the conditional expression, which is typically realized by invoking an SMT (satisfiability modulo theories) solver~\cite{SMT}, e.g., Z3~\cite{z3}. If the SMT solver cannot find a solution, we consider the corresponding program path as infeasible. Therefore, symbolic execution can be used to discover dead code. However, there may be little dead code in EVM bytecode, because the compiler can eliminate dead code during the compilation of smart contracts. To accelerate our analysis, we consider the conditional expression, which is equal to “0” as unsatisfiable and all other conditional expressions as satisfiable, without checking their satisfiability.

\begin{table}
	\scriptsize
	\caption{The Information Required to Detect Each Contract Defect}
	\label{tab:info}
	\begin{center}
		\begin{tabular}{l <{\centering} | p{50pt}  <{\centering} |p{40pt} <{\centering} }
			\hline
			\textbf{Contract Defect} & \textbf{Control Flow Information} & \textbf{Symbolic State} \\
			\hline
			\textit{Transaction State Dependency (TSD)}  &  & \checkmark  \\
			\hline
			\textit{DoS Under External Influence (DuEI)}  &  \checkmark  &  \checkmark  \\
			\hline
			\textit{Strict Balance Equality (SBE)}   &  \checkmark &  \checkmark  \\
			\hline
			 \textit{Reentrancy (RE)} &  \checkmark  &  \checkmark  \\
			\hline
			\textit{Nested Call (NC)}  &  \checkmark &  \checkmark   \\
			\hline
			\textit{Greedy Contract (GC)} & \checkmark  & \checkmark  \\
			\hline
			\textit{Unchecked External Calls (UEC)} & & \checkmark  \\
			\hline
			 \textit{Block Info Dependency (BID)} &  \checkmark  &  \\

		\end{tabular}
	\end{center}
\end{table}

\subsection{Feature Detector}
To detect contract defects at the bytecode level, we need to identify some specific behaviors from their opcodes. In this part, we introduce three features that we use when detecting contract defects. 

\subsubsection{Money Call}
To detect \textit{Reentrancy}, we need to identify whether a smart contract can transfer Ethers to other contracts. Ethereum provides three methods to transfer Ethers, i.e., \textit{address.send()}, \textit{address.transfer()}, \textit{address.call().value()}. All of these three methods generate a \textit{CALL} instruction. However, only detecting the \textit{CALL} instruction is not enough, as many other behaviors can also generate \textit{CALL} instruction, e.g.,  calling functions on other contracts or library. In this paper, if a \textit{CALL} instruction is generated by functions which are used to transfer Ethers, we call this \textit{CALL} instruction a \textit{Money-CALL}. Otherwise, the \textit{CALL} instruction is a \textit{No-Money-CALL}.  \textit{CALL} reads seven values from EVM stack. The first three values represent the gas limitation, recipient address, transfer amounts, respectively. If the transfer amount is larger than 0, the \textit{CALL} instruction is a \textit{Money-CALL}. 

However, only detecting \textit{Money-CALL} is still not enough, as \textit{address.send()} and \textit{address.transfer()} will limit the maximum gas consumption to 2300, which is not enough to send Ethers. Therefore, these two methods also cannot cause \textit{Reentrancy}. If the \textit{CALL} instruction is generated by \textit{address.send()} and \textit{address.transfer()}, a specific number ``2300" will be pushed into EVM stack, which represented the maximum gas consumption. So, if \textit{CALL} instruction reads a specific number ``2300" from the EVM stack, the \textit{CALL} instruction is generated by \textit{address.send()} and \textit{address.transfer()}. We call this \textit{CALL} instruction a \textit{Gas-Limited-Money-CALL}. Otherwise, if the first value read by \textit{CALL} instruction does not contain a specific value ``2300", we assume that the \textit{CALL} instruction is generated by \textit{address.call().value()}. We call this \textit{CALL} instruction a \textit{Gas-Unlimited-Money-CALL}. 

\subsubsection{Loop Block}
After constructing the CFG,  we need to detect which block is the start of a loop and which blocks make up the body of the loop. To detect this information,  we first traverse the path of the CFG by utilizing \textit{DFS} (Depth-first-search)~\cite{DFS} and then flag all blocks we visit. If there is a block that has been visited, this block is the start of a loop, and other blocks in this cycle are the loop bodies. Since some smart contracts are very complicated, it may contain a large number of paths. To reduce the computational effort, we use the strategy of pruning.  For example, block A is the destination of many other blocks, and we find the path of block A does not contain any cycles. We do not need to visit the remaining paths when other paths encounter block A.

\subsubsection{Payable Function}
A smart contract can receive Ethers only if it contains payable functions~\cite{Ethereum_yellow_paper}. To detect whether a function is payable or not, we can inspect the first block of each function. \textit{CALLVALUE} instruction is used to get the received Ether amount. If a smart contract receives Ethers, \textit{CALLVALUE} instruction will get a non-zero value. This value can be checked by the \textit{ISZERO} instruction to know whether a transaction contains Ethers. If the function is not payable, when receiving Ethers, it will throw an exception and terminate the execution.

To find the first block, we first rank all instructions by their index ID. All conditional jumps positioned before the first \textit{JUMPDEST} instruction are the start position of each function. EVM uses a hash value to identity functions; when EVM receives a function call, it first compares the received value to each function's hash value. If a function's hash value is equal to the received hash value, it will jump to the destination, which indicates a function's start position. Otherwise, it will fall to \textit{fallback} function, whose start position is the first \textit{JUMPDEST} instruction.

\subsection{DefectChecker}

Table~\ref{tab:info} describes the information required to detect each kind of contract defect. To detect TSD and UEC, \textit{DefectChecker} only needs symbolic states computed by symbolic execution, as we only need to check whether \textit{ORIGIN} and \textit{CALL} instructions are read by \textit{EQ} and \textit{ISZERO} instruction, respectively. \textit{DefectChecker} only needs control flow information to detect BID, as we only need to check whether the conditional expression contains block related instructions, e.g., "BLOCKHASH". 

To detect the other 5 contract defects, \textit{DefectChecker} needs both control flow information and symbolic states.  In the previous subsection, we introduce three features detected by the feature detector, i.e., \textit{Money Call}, \textit{Loop Block}, and \textit{Payable function}. \textit{Money Call} needs symbolic states, so to detect it, \textit{DefectChecker} needs check the values on the EVM stack. \textit{Loop Block} and \textit{Payable function} require control flow information, as they both need CFGs to locate the loop and the start of the function, respectively. \textit{NC}, \textit{DuEI}, \textit{GC}, and \textit{Reentrancy} all need to detect \textit{Money Call}. \textit{DuEI} and \textit{NC} also need to detect  \textit{Loop Block}; \textit{GC} needs to detect \textit{Payable function}. To detect \textit{Reentrancy}, \textit{DefectChecker} needs to travel all the paths that contain the  \textit{Gas-Unlimited-Money Call}, which needs the help of the CFG. To detect \textit{SBE}, \textit{DefectChecker} needs to check whether the \textit{BALANCE} instruction is read by the \textit{EQ} instruction in the conditional expressions, which needs both control flow information and symbolic state. 

Below we describe the detailed patterns that we use to determine whether a smart contract contains one or more of the contract defects.

\subsubsection{Transaction State Dependency} \textit{tx.origin} generates an \textit{ORIGIN} instruction. We first locate all \textit{ORIGIN} instructions. We then check whether there is an  \textit{ORIGIN} that is read by an \textit{EQ} instruction. The \textit{EQ} instruction reads two values from EVM stack and verifies whether these two values are equal. If the contract contains this kind of contract defect \textit{ORIGIN} instruction will compare to an address value. Ethereum uses a 40-bit value to indicate an address, and all addresses conform to the EIP55 standard~\cite{EIP55}.

\subsubsection{DoS Under External Influence} If a smart contract contains this contract defect, there will be a part of the instructions that check the return value of the \textit{Money CALL}, and then terminate the loop. To detect this contract defect, we first find loop-related blocks. Then, we check whether there is a block that contains \textit{Money CALL}, and the type of the block is \textit{conditional}, as it needs to check the return value. Then, this block jumps to a block, which type is \textit{terminal}.

\subsubsection{Strict Balance Equality} This kind of contract defect can make a part of the code never be executed. We need to check whether there is a conditional expression that contains the related pattern. \textit{BALANCE} instruction is used to get the balance of a contract. If a \textit{BALANCE} instruction is read by \textit{EQ}, it means there is a strict balance equality check. If this check happens at a conditional jump expression, it means this contract contains this contract defect.

\subsubsection{Reentrancy}
The \textit{SLOAD} instruction is used to get a value from storage~\cite{Ethereum_yellow_paper}. It reads a value (named \textit{Slot ID}) from the EVM stack and puts the result that reads from storage back onto the EVM stack. Using \textit{listing 4} as an example, \textit{Victim} contracts do not make the balance of an \textit{Attack} contract to zero (L7) before sending Ethers (L6), which allows an \textit{Attack} contract to withdraw Ethers again. To detect this contract defect, we first need to obtain paths that contain \textit{Gas-Unlimited-Money-Call}, because only this kind of \textit{CALL} can cause \textit{Reentrancy}.
We then need to obtain all conditional expressions on these paths. The amount that is sent by the victim contract is usually checked before sending it to attacker contracts, and this amount is loaded from \textit{storage}. In this case, we need to check if the conditional expression contains \textit{SLOAD} instructions and get its \textit{Slot ID}. If this value still holds and does not be updated when executing \textit{CALL} instruction, it means \textit{CALL} instruction can be executed again and cause \textit{Reentrancy}. To check whether the storage value is updated, we need to detect whether the same \textit{Slot ID} that is read by \textit{SLOAD} is written by \textit{SSTORE} instruction. (\textit{SSTORE} instruction is used to save data to memory. It reads two values from EVM stack, i.e., slot id and value that are written to storage.)


\subsubsection{Nested Call} 
\label{sec:nc}
Using \textit{listing 5} as an example, array \textit{members} is a storage variable, all of its value, including its length,  are stored on storage. To get its length, \textit{SLOAD} instruction reads its \textit{Slot ID} $\delta$  from EVM stack, and this value is the position that stores the value of \textit{members.length}. To detect this contract defect, the first step is to find the start block of a loop and get the \textit{Slot ID}. Then, we need to check whether this loop limits its size. If the loop limits its size, the same \textit{Slot ID} $\delta$ will be read in the loop body again, and this value will be compared with another value. If a smart contract contains a loop that does not limit its size but contains a \textit{Money-Call}, \textit{Nest Call} is detected in this contract.

\subsubsection{Greedy Contract} A smart contract can transfer money through a \textit{Money CALL} or \textit{selfdestruct} function. \textit{selfdestruct} function generates \textit{SELFDESTRUCT} instruction. If a smart contract contains payable functions but does not have either a \textit{Money CALL} or \textit{SELFDESTRUCT} instruction, the contract is a \textit{Greedy Contract}.

\subsubsection{Unchecked External Calls} The external call returns a boolean value. If the result is checked by the contract, it will generate an \textit{ISZERO} instruction. To detect this contract defect, we first locate \textit{CALL} instructions. Then, we check whether each \textit{CALL} instruction is read by \textit{ISZERO}. If there is a \textit{CALL} that is not checked by \textit{ISZERO}, this contract defect is detected.

\subsubsection{Block Info Dependency} Detecting this contract defect is similar to \textit{Strict Balance Equality}. This contract defect can allow miners to control the contract, as miners can change the value of some block information, which affects the result of the conditional expression. If the conditional expression contains block related instructions, i.e., \textit{``BLOCKHASH", ``COINBASE", ``NUMBER", ``DIFFICULTY", ``GASLIMIT"}, it means the contract contains this contract defect.

	\section{Evaluation}

To measure the efficacy of \textit{DefectChecker}, we present results based on applying it to an open-sourced dataset and present our experimental results analysis in this section.

\subsection{Experimental Setup}
All experiments were performed on a PC running Mac OS 10.14.4 and equipped with an Intel i7 6-core CPU and 16 GB of memory. We use Solidity 0.4.25 as the compiler to compile source code into bytecode, and use EVM 1.8.14 to disassemble the bytecode to its opcodes.

\subsection{Dataset}

\begin{table}
	\scriptsize
	\caption{Some Features of Dataset }
	\label{tab:dataset}
	\resizebox{0.5\textwidth}{!}{%
	\begin{tabular}{l | r | r | r | r }
	\hline
    \textbf{Features} & \textbf{Min} & \textbf{Max} & \textbf{Mean} & \textbf{SD} \\
    \hline
    Lines of Code &5&2,239&393.6& 356.8\\
    \hline
    \# of Functions& 1 & 174 & 30.1 & 621.6\\
    \hline
    \# of Instructions & 7 & 15,355 & 3,597.3 & 2,523.7\\
    \hline
    CC & 1 & 132 & 30.3 & 22.4\\
     \hline
    Ethers & 0 & 1,500,000 & 7,844.9 & 1,704,552.7\\
    
	\end{tabular} }
\end{table}

The dataset we used to evaluate \textit{DefectChecker} was released in our previous work~\cite{fse-smell}. We first crawled all 17,013 open sourced smart contracts from Etherscan. Then, we randomly selected 600 smart contracts from these contracts. We found 13 smart contracts do not contain any contents. Thus, we removed them from our dataset. Finally, we obtained 587 smart contracts from Etherscan. These contracts have 231,098 lines of the code and more than 4 million Ethers in their balance. 

Table~\ref{tab:dataset} shows some key features of the dataset, i.e., lines of code,  number of functions in the contracts, number of instructions in the contracts, cyclomatic complexity~\cite{complexity} and Ethers hold by the contracts. Cyclomatic complexity is a software metric that indicates the complexity of a program, and it is computed by analyzing the control flow graph. The formulation to compute it is: \textit{\textbf{E - N + 2P}}.   E is the number of edges on CFG; N is the number of nodes on CFG and P is the number of connected components on CFG. Since CFG is a connected graph, so P always equal to 1, and the formulation can be simplified as: \textit{\textbf{E - N + 2}}. 

The simplest contract in our dataset only contains one constructor function with 7 instructions and a cyclomatic complexity of 1. The contract with the highest cyclomatic complexity has 11,696 instructions and 2,004 lines of code. The richest contract in our dataset holds 1.5 million Ethers, while the poorest contract has no Ethers in its balance. 

Two authors of our previous work manually labeled the dataset. They both have three years of experience working on smart-contract-based development and research, and took part in the process of defining contract defects. Thus, they have a very good understanding of the smart contract programming and contract defects introduced in this paper. They first manually labeled the dataset independently. Then, they discussed the disagreements after completing the labeling process and gave the final results. Their overall Kappa﻿ value~\cite{kappa} was 0.71, which shows a substantial agreement between them. 

In this work, we developed a tool named \textit{DefectChecker} to detect eight contract defects with severity impact levels 1-3. The numbers of each type of contract defect in our dataset are shown in Table~\ref{tab:res}. This shows that \textit{Block Info Dependency} is the most frequent contract defect in our dataset, while \textit{Transaction State Dependency} and \textit{Strict Balance Equality} are the least popular. Their numbers are 42, 5, and 5, respectively.  \textit{DefectChecker} aims at Solidity version 0.4.0+, which is the most widely used version at the time of writing this paper~\cite{EtherScan}. However, some smart contracts are designed for Solidity version 0.2.0+ and 0.3.0+. Thus, we removed eight smart contracts and used the remaining 579 smart contracts as our ground truth.

Among the six tools we introduced in Table~\ref{tab:input}, only \textit{Zeus} open sourced their dataset. However, \textit{Zeus} still has four kinds of defects which are not included in their dataset. Also, the \textit{Zeus} authors did not provide the detail of how to built their dataset. Their paper only mentioned that ``they manually validated each result” without providing any details, e.g., the number of people who labeled the dataset, and whether they are professional smart contract developers or not. Thus, we did not use these datasets.

\subsection{Evaluation Methods and Metrics}

There are seven measurements obtained from our experiments: True Positive (TP), True Negative (TN), False Positive (FP), False Negative (FN), Precision (P), Recall (R) and F-Measure (F). TP indicates the results which correctly predict a contract defect in a smart contract. TN indicates the results which correctly predict a smart contract does not have a defect. FP and FN indicate the results which incorrectly predict that a smart contract contains and does not contain a contract defect. $\mathit{Precision}$, $\mathit{Recall}$,  and $\mathit{F\mhyphen Measure}$ can be calculated as: 

\begin{eqnarray}
\mathit{Precision} = \frac{\#TP}{\#TP + \#FP} \times 100\%
\end{eqnarray}

\begin{eqnarray}
\mathit{Recall} = \frac{\#TP}{\#TP + \#FN} \times 100\%
\end{eqnarray}

\begin{eqnarray}
\mathit{F\mhyphen Measure}= \frac{2 \times Precision \times Recall}{ Precision + Recall} \times 100\%
\end{eqnarray}

\begin{table}\centering
	\normalsize
	\caption{Experimental results for \textit{DefectChecker}.}
	\label{tab:res}

	\resizebox{0.5\textwidth}{!}{%
	\begin{tabular}{l || r|r|r|r|r|  r|  r|  r }
		\hline
		\textbf{Defects} & \textbf{\#Defects} & \textbf{\#TP}  & \textbf{\#TN}& \textbf{\#FP} & \textbf{\#FN} &  \textbf{P(\%)}  &   \textbf{R(\%)} &   \textbf{F(\%)}  \\
		\hline
		\textit{TSD}  & 5 & 5& 474& 0 & 0 & 100.0 & 100.0 & 100.0\\
		\hline
		\textit{DuEI}  & 6 & 6& 466& 7 & 0 & 46.2 & 100.0 & 63.2\\
		\hline
		\textit{SBE}   & 5 & 4 &474& 0  & 1 & 100.0 &80.0&88.9 \\
		\hline
		\textit{RE} & 12 & 10  & 461 & 6 & 2 & 62.5 & 83.3 & 71.4\\
		\hline
		\textit{NC}  & 13 & 9 & 464& 2 & 4 & 81.8 & 69.2 & 75.0\\
		\hline
		\textit{GC} & 6 & 6  & 473& 0 & 0 & 100.0 & 100.0& 100.0 \\
		\hline
		\textit{UEC} & 22 & 20 & 454 & 3 & 2 & 87.0 & 90.9 & 88.9\\
		\hline
		\textit{BID}  & 42 & 41 & 437 & 0 & 1 & 100.0 & 97.6& 98.8\\
		
	\end{tabular} } 
	
\end{table}

\subsection{Experimental Results and Analysis}

Table~\ref{tab:res} summarizes the results of applying \textit{DefectChecker} to our previous work's dataset. The first column is the contract defects that need to be detected. The second column is the number of contract defects in our dataset (ground truth). The remaining seven columns are used to measure the performance of \textit{DefectChecker}. Below, we discuss the analysis of each contract defect.

\textbf{\textit{(1). Transaction State Dependency.}} \textit{DefectChecker} detects 5 smart contracts containing this contract defect among 579 smart contracts with 0 false positives and negatives.

\textbf{\textit{(2). DoS Under External Influence.}} \textit{DefectChecker} detects 13 smart contracts that have this contract defect among 579 smart contracts with 7 false positives and 0 negatives. The 7 errors are due to the error identification of a loop. 

In our detection method, we first split the bytecode into several blocks. Then, symbolic execution is used to find the edge between blocks. We traverse the path of CFG by using \textit{DFS}. If there is a block that has been visited, we regard this block as the start of the loop (See Section 3.4.2). Since we regard all the paths are reachable, thus we only flag whether two blocks have an edge. This mechanism leads to false positives in detecting loops. 

In Listing 10, all the L9, L10, and L11 hold a single block, respectively, and function \textit{sub()} holds several blocks. EVM first executes the block of line 9, then executes the blocks of function \textit{sub()} in line 2. After the execution of blocks of line 10, line 11, respectively, the blocks of function \textit{sub()} will be executed again. Therefore, when traversing the CFG by using DFS, we can find that there is a cycle (fun sub()$\rightarrow$L10$\rightarrow$L11 $\rightarrow$fun sub()). Since we regard all the paths are reachable, we cannot know that the blocks of function \textit{sub()} cannot jump the  block of L10, after executing the block of L11. 

This kind of false positive can be addressed if we execute the loop continuously. Using a loop ``\textit{for(int i = 0; i \textless  100; i++)}” as an example; we need to record the state of variable \textit{i}, and check whether the expression \textit{(i \textless 100}) is satisfied or not. If we prove the loop can execute continuously, we can confirm it is a real loop not the error we show in Listing 10. However, we need the assistance of an SMT solver to execute the loop, and executing the loop continuously is also time consuming. Thus, we believe the advantages of removing the use of an SMT solver in our approach outweighs the disadvantages. 



\begin{lstlisting}[caption={Error Loop Example}]
library SafeMath {
  function sub(uint256 a, uint256 b) internal returns (uint256) {
	assert(b <= a);
	return a - b;}}
contract Mainsale {
  using SafeMath for uint256;
  uint256 public total;
  function() payable {
	uint amount = total.sub(100);
	msg.sender.transfer(amount);
	uint contri = msg.value.sub(amount);}}
\end{lstlisting}\vspace{-0.1cm}

\textbf{\textit{(3). Strict Balance Equality.}} \textit{DefectChecker} detects 4 smart contracts that contain \textit{Strict Balance Equality} with 0 false positives and 1 false negative. The cause of the error is that the contract defect related to several functions. For example, the contract in listing 11 uses a global variable \textit{balance} to represent the contract's balance. Callers first call function \textit{getBalance} to obtain the balance. The balance will then be checked in Line 5. To detect this contract defect, we need to know that the global variable \textit{balance} represents the contract balance.  Therefore, the contract defect can only be detected when we know users will first invoke \textit{getBalance()} and then call \textit{DefectFunction()}.  However, it is not easy to detect this contract defect at the bytecode level, as the two operations (i.e., \textit{balance == 1 eth} and \textit{balance = this.balance}) are in two independent functions, and we do not know the calling sequence.  

\begin{lstlisting}[caption={Strict Balance Equality - False Negative Example}]
contract Demo{
  uint balance = 0;
  function getBalance(){ balance = this.balance;}
  function DefectFunction() {
	if(balance == 1 eth)
	doSomthing;} }
\end{lstlisting}

\textbf{\textit{(4). Reentrancy.}} \textit{DefectChecker} detects 16 smart contracts that contain \textit{Reentrancy}, with 6 false positives and 2 false negatives. The false positives are because of error-money-call detection. A smart contract contains \textit{Reentrancy} must have a \textit{Gas-Unlimited-Money-Call}. To detect it, we first need to check whether the gas limits set are larger than 2,300 gas and the transfer amount is larger than 0. However, in some examples, these two values are represented by complicated symbolic expressions. Some expressions also contain values that read from storage (read by \textit{SLOAD}). Thus their specific values can not be determined by static analysis. Therefore, \textit{DefectChecker}  failed to detect them. When \textit{DefectChecker} encounters complicated symbolic expressions, the default value is larger than 2,300 gas and larger than 0, this leads to false positives. When detecting this contract defect, we need to check whether the Slot ID read by \textit{SLOAD} instruction still holds when executing \textit{CALL} instruction. Some Slot IDs are also represented by complicated symbolic expressions. \textit{DefectChecker} failed to detect whether they are equal, which leads to reporting false negatives.

When detecting Money-Call, we use \textit{Gas-Limited-Money-Call} as default, if we cannot figure out the exact value of the gas limit symbolically. We also conduct another experiment, which uses Gas-Limited-Money-Call as the default. However, \textit{DefectChecker}  failed to detect any \textit{Reentrancy} default. The reason is that the \textit{Gas-Limited-Money-Call} usually is easy to detect, as \textit{address.transfer(), address.send()} will put a specific value ``2300" to the EVM stack. Thus, we just need to detect the specific value. However, the gas limit of Gas-Unlimited-Call is not easy to detect, as it usually uses a complicated expression to represent the gas. Since \textit{address.call.value()} will not change the gas cost. In most situations, this method will not lead to an out-of-gas error. This is the reason why we use Gas-Unlimited-Call as our default. 

\textbf{\textit{(5). Nested Call.}}  \textit{DefectChecker} detects that 11 smart contracts contain a \textit{Nested Call} defect. Among these 11 smart contracts, we have 2 false positives and 4 false negatives. The cause of the false positives is also the error identification of the loop, which is the same with \textit{DoS Under External Influence}. The false negatives are because of the complicated data structure. When detecting this contract defect, the first step is to know whether the loop iterations are related to the array's length. We use the \textit{SLOAD} instruction related pattern to obtain the loop iterations, as described in Section~\ref{sec:nc}. However, as shown in Listing 12, \textit{self} is a structure and its length is obtained through an external function. Since external functions can be designed in different ways, it is challenging to design a pattern to detect it.

\begin{lstlisting}[caption={Nest Call - False Negative Example}]
for (uint i; i<self.keys.length; i++) {
	self.data[ self.keys[i]].transfer(1 Ether);}
\end{lstlisting}

\textbf{\textit{(6). Greedy Contract.}}  \textit{DefectChecker} detects 6 \textit{Greedy Contracts}, with 0 false positives and negatives.

\begin{lstlisting}[caption={Unchecked External Call - False Positive Example}]
function Example(Address addr) returns (bool) {
	return addr.send();}
\end{lstlisting}

\textbf{\textit{(7). Unchecked External Call.}}  \textit{DefectChecker} reports 23 contracts have this kind of contract defect, with 3 false positives and 2 false negatives. We analyzed the false positive examples and find that these contracts use the return value of \textit{send()} as function's return value and check the return value in other functions. For example, \textit{addr.send()} as shown in listing 13 is the return value of function \textit{Example}, and the value is checked in the callee programs. The false negatives are because the defect happens in a constructor function, while the bytecode of the constructor function is not contained in runtime bytecode. Therefore, we missed it. However, the contract defects in the constructor function will not harm the deployed contracts, as the constructor function will only be executed once when deploying the contracts to the blockchain. 

\textbf{\textit{(8). Block Info Dependency.}} \textit{DefectChecker} detects 41 smart contracts contain this contract defects, with 0 false positives and 1 false negative. The cause of the false negative is similar to the one with \textit{Strict Balance Equality}. The defect contract uses a global variable to represent block information and uses this global variable in other functions, which causes the contract defect to be detected. 

\begin{table*}
	\caption{Input and Defects Detected of Each Tool}
	\label{tab:input}
	\centering
	\begin{tabular}{l || l | c | c | c | c | c | c | c | c | c}
		\hline
		\textbf{Tools} & \textbf{Input} & \textbf{TSD} & \textbf{DuEI} & \textbf{SBE} & \textbf{RE} & \textbf{NC} & \textbf{GC} & \textbf{UEC} & \textbf{BID} & \textbf{\# of Other Defects}\\
		\hline
		\textit{DefectChecker}  & Bytecode &  \checkmark &  \checkmark&  \checkmark&  \checkmark&  \checkmark&  \checkmark&  \checkmark&  \checkmark & 0\\
		\hline
		\textit{Oyente~\cite{oyente}}  & Bytecode &  &  & &  \checkmark& &  &  \checkmark&  \checkmark & 1\\ 
		\hline
		\textit{Maian~\cite{Maian}}  & Bytecode &&&&&&\checkmark&& &2\\
		\hline
		\textit{Securify~\cite{Securify}}  & Bytecode &&&&\checkmark&&&\checkmark& & 7\\
		\hline	
		\textit{Mythril~\cite{Mythril}}  & Bytecode &  \checkmark &  \checkmark&  \checkmark&  \checkmark&  \checkmark& &  \checkmark& &28\\
		\hline
		\textit{Contractfuzzer~\cite{Contractfuzzer}}  & Bytecode + ABI &  &  &  &  \checkmark&  & &  \checkmark&  \checkmark & 3\\
		\hline
		\textit{Zeus~\cite{Zeus}}  & Source Code &  \checkmark &  &  & \checkmark &  &  &  \checkmark&\checkmark & 3\\
	\end{tabular}  \vspace{-0.3cm}
\end{table*}

\begin{table}
	\scriptsize
	
	\caption{Experiment result of \textit{Oyente}.} \vspace{-0.2cm}
	\label{tab:oyente}
	\centering
		\resizebox{0.5\textwidth}{!}{%
		\begin{tabular}{l | |r|r|r|r|r |r |r|r}
		\hline
		\textbf{Defects} & \textbf{\#Defects} & \textbf{\#TP}  & \textbf{\#TN}& \textbf{\#FP} & \textbf{\#FN} &  \textbf{P(\%)}  &   \textbf{R(\%)} &   \textbf{F(\%)}  \\
		\hline
		\textit{RE} & 12& 2& 94 & 373 & 10 & 2.1& 16.7 & 3.7\\
		\hline
		\textit{UEC} & 22&16 & 448 & 9 & 6 & 64.0 & 72.7 & 68.1\\
		\hline
		\textit{BID}  & 42 & 11 & 431& 6 & 31 & 64.7& 26.2 & 37.3 \\
		
	\end{tabular} }
	
\end{table}

\begin{table}
	\scriptsize
	
	\caption{Experiment result of \textit{Mythril}.} \vspace{-0.2cm}
	\label{tab:Mythril}
	\centering
\resizebox{0.5\textwidth}{!}{%
	\begin{tabular}{l | |r|r|r|r|r |r |r|r}
			\hline
			\textbf{Defects} & \textbf{\#Defects} & \textbf{\#TP}  & \textbf{\#TN}& \textbf{\#FP} & \textbf{\#FN} &  \textbf{P(\%)}  &   \textbf{R(\%)} &   \textbf{F(\%)}  \\
		\hline
		\textit{TSD}  & 5 &0 & 474& 0 & 5 & 0 & 0 & 0 \\
		\hline
		\textit{DuEI}  & 6 & 1 & 245 & 228 & 5 & 0.4 & 16.7 & 0.8 \\
		\hline
		\textit{SBE}   & 5 & 0& 474& 0 & 5 & 0 &0 &0 \\
		\hline
		\textit{RE} & 12 & 5& 280&  187 & 7 & 2.6 & 41.7& 4.9\\
		\hline
		\textit{NC}  & 13 & 2&414& 52 & 11 & 3.7& 15.4& 6.0\\
		\hline
		\textit{UEC} & 22 & 11 & 436 & 21 & 11 & 34.4 & 50.0 & 40.8\\
		
	\end{tabular} }
	
\end{table}

\begin{table}[htb]
	\scriptsize
	\caption{Experiment result of \textit{Securify}.} \vspace{-0.2cm}
	\label{tab:security}
	\centering
	\resizebox{0.5\textwidth}{!}{%
	\begin{tabular}{l | |r|r|r|r|r |r |r|r}
		\hline
		\textbf{Defects} & \textbf{\#Defects} & \textbf{\#TP}  & \textbf{\#TN}& \textbf{\#FP} & \textbf{\#FN} &  \textbf{P(\%)}  &   \textbf{R(\%)} &   \textbf{F(\%)}  \\
		\hline
		\textit{RE} & 12 & 1& 439& 28 & 11 &  3.5& 8.3 & 4.9\\
		\hline
		\textit{UEC} & 22 & 10 & 457& 0 & 12 & 100.0& 45.5 & 62.5\\
		
	\end{tabular} }
\end{table}

\subsection{Comparison with state-of-the-art tools}
In our previous work, we investigated whether there are existing tools that can detect some of the contract defects we have defined. We first collected all the papers from top Security and SE conferences/journals, i.e., CCS, S\&P, USENIX Security, NDSS, ACSAC, ASE, FSE, ICSE, TSE, TIFS, and TOSEM from 2016 to 2019. Then, we only retain the papers whose titles have the key words ``smart contract", ``Ethereum" or ``blockchain". After that, we manually read the abstract to verify their relevance. Finally, we found only four papers that are related to smart contract defects, i.e., Oyente~\cite{oyente}, Maian~\cite{Maian}, Zeus~\cite{Zeus}, and ContractFuzzer~\cite{Contractfuzzer}. 

To enlarge our baseline methods, we use the same method as proposed by Kitchenham et al.~\cite{kitchenham2004procedures}. We first read the references of these 4 relevant papers, and tried to find whether there are existing tools that can detect the defined contract defects. If there is a relevant paper, we read its references repeatedly, until no new paper can be found. In this way we also found two other tools, i.e., Securify~\cite{Securify} and Mythril~\cite{Mythril}. 
 
Table~\ref{tab:input} shows the input and contract defects that can be detected by these tools. The last column shows the number of the defects can be detected by these tools except the mentioned 8 contract defects.  As we know, the bytecode of smart contract on Ethereum are visible to everyone, but only less than 1\% of the smart contracts open up their source code~\cite{TokenScope}. Therefore, detecting contract defects from the bytecode level is very important. To make the comparison fair, we select \textit{Oyente, MAIAN, Securify} and \textit{Mythril} as our baseline tools, since they can detect contract defects at the bytecode level, the same as \textit{DefectChecker}.  However, we found that \textit{Maian} has not been updated to support the latest Ethereum environment and so we could not run \textit{MAIAN} on our dataset. For example, they use methods provided by \textit{web3}~\cite{Web3py} to obtain contracts' information on Ethereum. However, the methods they used have been removed and did not support the current version of Ethereum that we used. In addition, \textit{DefectChecker} gets 100\% F-Measure when detecting \textit{Greedy Contract}. In this case, we do not compare with \textit{MAIAN}, and choose \textit{Oyente, Securify} and \textit{Mythril} as our baseline tools.

\begin{table*}[htb]
	\caption{Result Comparison(F-Measure) between Four Tools}
	\label{tab:all}
	\centering
		\resizebox{0.7\textwidth}{!}{%
	\begin{tabular}{c ||r|r|r|r|r |r |r|r}
		\hline
		\textbf{Tools} &  \textbf{TSD} & \textbf{DuEI} & \textbf{SBE} & \textbf{RE} & \textbf{NC} & \textbf{GC} & \textbf{UEC} & \textbf{BID} \\
		\hline
		\textit{DefectChecker}  & \textbf{100.0\%} &  \textbf{63.2\% }&  \textbf{88.9\%} &  \textbf{71.4\% }&  \textbf{75.0\%}&  \textbf{100.0\%}&  \textbf{88.9\%}&  \textbf{98.8\% }\\
		\hline
		\textit{Oyente}   & / & / &/ &  3.7\% & /& / &  68.1\%&  37.3\% \\ 
		\hline
		\textit{Securify}  &/&/&/&4.9\%&/&/&62.5\%&/\\
		\hline	
		\textit{Mythril}   &  0\% &  0.8\%&  0\%&  4.9\%&  6.0\%&/ &  40.8\%& /\\

	\end{tabular}}  
\end{table*}

\textit{Oyente} detects three kinds of security-related vulnerabilities for smart contracts. These three kinds of security-related vulnerabilities are the same as our \textit{Unchecked External Calls}, \textit{Block Info Dependency} and \textit{Reentrancy}.  \textit{Mythril}~\cite{Mythril} is a tool developed by \textit{ConsenSys}, which is a leading global blockchain technology company. They find security problems from online posts or news, which is similar to our previous work~\cite{fse-smell}. Our previous work analyzed the posts from StackExchange posts and defined 20 contract defects. \textit{Mythril} can detect 6 contract defects as shown in Table~\ref{tab:Mythril}. \textit{Securify} is a smart contract security analyzer that takes EVM bytecode as input. It first decompiles EVM bytecode and analyzes the semantic facts of the decompiled code. In our study, \textit{Securify} uses several security patterns to detect related vulnerabilities. \textit{Securify} can detect \textit{Reentrancy} and \textit{Unchecked External Call}, which can also be detected by \textit{DefectChecker}. 

Table~\ref{tab:oyente} shows the results of running \textit{Oyente} on our previous dataset~\cite{fse-smell}. The F-score of \textit{Oyente} in detecting RE, UEC, and BID are 3.7\%, 68.1\%, and 37.3\%, respectively, while the numbers for DefectChecker are 71.4\%, 88.9\%, and 98.8\%, respectively. We found that \textit{Oyente} only considers \textit{BLOCKHASH} instructions when detecting \textit{Block Info Dependency}, while there are many other instructions, e.g. NUMBER (NUMBER instruction is used to get block's number),  that can lead to this contract defect. Besides, \textit{Oyente} also has many false positives when detecting \textit{Reentrancy}. The reason is that they do not distinguish between \textit{send()}, \textit{transfer()} and \textit{call()} functions at the bytecode level, while \textit{send()} and \textit{transfer()} will limit gas to 2300 unit, which cannot cause \textit{Reentrancy}. In addition, the most important reason for these errors is \textit{code coverage}. Code coverage means the percentage of instructions executed. The average code coverage for \textit{Oyente} is 18.9\%, while the  number for \textit{DefectChecker} is 77.1\%. Low code coverage means only a small part of the code can be analyzed for contract defect occurrence, which can lead to a large number of false positives and negatives. There are three reasons that lead to the low coverage of Oyente compared to DefectChecker. First, \textit{Oyente} checks whether a path can be reached, while \textit{DefectChecker} assumes that all the paths are reachable. \textit{Oyente} also only optimizes for Solidity Version 0.4.19, but there is a wide version coverage in our dataset. Finally, the jump positions of some unconditional jump might not be easy to find. To be specific, the jump position might be a result of a complicated expression. Thus both \textit{Oyente} and \textit{DefectChecker} can fail to detect these unconditional jumps, and it is the reason why DefectChecker misses some blocks.

%
%

Table~\ref{tab:Mythril} shows the results of \textit{Mythril}. \textit{Mythril} fails to detect \textit{Transaction State Dependency} and \textit{Strict Balance Equality} in our dataset. In addition, its results contain many false positives, especially in detecting \textit{Reentrancy} and \textit{DoS Under External Influence}. We found that \textit{Mythril} is similar to \textit{Oyente} - it fails to distinguish between \textit{call()} with \textit{transfer()} and \textit{send()}, which will not lead to \textit{Reentrancy}. Besides, \textit{Mythril} failed to distinguish loop related patterns, which lead to errors when detecting loop related defects, e.g., \textit{DoS Under External Influence} or \textit{Nest Call}. 

Table~\ref{tab:security} presents the results of \textit{Securify}. \textit{Securify} can detect two common defects with \textit{DefectChecker}, i.e., \textit{Reentrancy} and \textit{Unchecked External Call}. All the \textit{DefectChecker}, \textit{Oyente}, \textit{Mythril}, and \textit{Securify} can detect these two defects. The performance of \textit{Securify} in testing \textit{Reentrancy} (4.9\%) is better than \textit{Oyente} (3.7\%), and similar to \textit{Mythril} (4.9\%), but much worse than \textit{DefectChecker} (71.4\%). In terms of detecting \textit{Unchecked External Call}, the F-score of \textit{Securify} (62.5\%) is a little bit worse than \textit{Oyente} (68.1\%) and much better than \textit{Mythril}. \textit{DefectChecker} still get the best F-score, which receives 88.9\% in detecting \textit{Unchecked External Call} 

To compare the results between all four tools, we add a comparison of F-measure in Table~\ref{tab:all}, which shows that \textit{DefectChecker} obtains the best F-measure of all four tools. 

\begin{table}[htb]
	\caption{Overall Precision, Recall, and F-Measure of Each Tool}
	\label{tab:overall}
	\centering
	\resizebox{0.45\textwidth}{!}{%
	\begin{tabular}{l || c | c | c }
		\hline
		\textbf{Tools} &  \textbf{O. P. (\%)} & \textbf{O. R (\%)} & \textbf{O. F. (\%) }  \\
		\hline
		\textit{DefectChecker}  & 88.3 & 90.9 & 88.8 \\
		\hline
		\textit{Oyente}   & 54.6  & 38.2 & 40.9 \\ 
		\hline
		\textit{Securify}  & 65.9& 32.4 & 42.2\\
		\hline	
		\textit{Mythril}   & 13.3 & 30.2 & 16.5\\
	\end{tabular}  }
\end{table}

\begin{table}[htb]
	\caption{Time Consumption of Each Tool}
	\label{tab:times}
	\centering
	\resizebox{0.45\textwidth}{!}{%
	\begin{tabular}{l || r | r | r | r }
		\hline
		\textbf{Tools} &  \textbf{Avg.} & \textbf{Max} & \textbf{Min}  & \textbf{S.D.} \\
		\hline
		\textit{DefectChecker}  & 0.15s& 2.42s & 0.04s & 5.43 \\
		\hline
		\textit{Oyente}   & 18.48s  & 1,096.32s & 0.28s & 2,877.64\\ 
		\hline
		\textit{Securify}  & 21.55s& 1,203.99s & 0.37s & 3,384.39\\
		\hline	
		\textit{Mythril}   &   103.55s & 2,480.26s & 1.58s & 13,063.80\\
	\end{tabular} }
\end{table}

We also calculate the overall precision, recall, and F-measure of all four tools on the whole experimental dataset. Using overall-precision as the example, the overall result is calculated by  $\frac{\sum_{i=1}^{n} p_{c_i} \times |c_i|  }{\sum_{i=1}^{n} |c_i| }  $, in which $p_{c_i}$ is the precision of the contract defect $i$, $|c_i|$ is the number of contract defect $i$ in the whole dataset. The results are given in Table~\ref{tab:overall}, which clearly shows that \textit{DefectChecker} obtains the best results in detecting contract defects. 

\noindent {\bf Time Consumption. } We calculate the time to analyze one smart contract to evaluate each tool. To make the evaluation accurate, we kill all the background processes in our machine when testing the tool to ensure the environment is clean. For each tool, we run it for 10 times and record the average time to test one smart contract in our dataset.

Table~\ref{tab:times} shows the time consumption results of each tool. The second column of the table gives the average time consumption to test a smart contract for each tool. The speed of \textit{DefectChecker} is the fastest in these four tools. It only needs 0.15s to analyze one smart contract. \textit{Oyente} and \textit{Securify} have similar running times. \textit{Oyente} needs 18.48s to analyze one smart contract, and the time for \textit{Securify} is 21.55s. \textit{Mythril} is the slowest tool; it needs 103.55s to analyze one smart contract. The maximum time to analyze a smart contract of \textit{DefectChecker} is 2.42s, while the time for \textit{Oyente}, \textit{Securify}, and \textit{Mythril} are 1096.32s, 1203.99s and 2480.26s, respectively. The simplest smart contract in our dataset only contains 7 lines with a single constructor function. \textit{DefectChecker} needs 0.04s to analyze it, while the time for \textit{Oyente}, \textit{Securify}, and \textit{Mythril} are 0.28s, 0.37s and 1.58s, respectively. \textit{DefectChecker} also has the smallest Standard Deviation value among these four tools, which shows that \textit{DefectChecker} has the most stable speed in analyzing a smart contract. 

In conclusion, the efficiency of these four tools is in order: \textit{DefectChecker} \textgreater  \textit{Oyente} \textgreater  \textit{Securify} \textgreater  \textit{Mythril}.

\subsection{Threats to Validity}
\textbf{Internal Validity.}  We used a dataset released in our previous work~\cite{fse-smell} as the ground truth to evaluate \textit{DefectChecker}. Since the people who developed \textit{DefectChecker} are the same as the people who labeled the dataset, it is likely that their familiarity with the dataset might lead to﻿ potential optimization or omissions when developing \textit{DefectChecker}. We tried to use the datasets of the baseline tools to evaluate \textit{DefectChecker}. However, we failed to find the dataset. Luu et al. run \textit{Oyente} on 19,366 contracts. They only manually check the correctness of some examples, instead of using a complete dataset to evaluate \textit{Oyente}. We can only find some false positive and true positive values on their paper. \textit{Securify} uses a complete dataset which consists of 100 smart contracts. However, they do not open their dataset to the public. \textit{Mythril} is a tool from industry. They even do not have an evaluation section in their technical papers. Thus, we had to build our own dataset. To reduce the influence of our dataset, we first wrote a few demo smart contracts when developing \textit{DefectChecker} and used these to conduct small-scale testing of our proposed tool. Then, we conducted large-scale testing by using real world bytecode we crawled from the Ethereum blockchain. The dataset is the same as that we introduced in Section 5. During this large-scale testing, we randomly choose a set of smart contracts that can find their source code. We use these smart contracts to improve the performance and patterns that are used to detect contract defects. We admit that the familiarity with the ground truth dataset might lead to a bias, but the methods we used to develop \textit{DefectChecker} can reduce this influence.

\textbf{External Validity.} The dataset we used to evaluate \textit{DefectChecker} is based on manual analysis, which may contain false positives and negatives. To address this problem, we double-checked the results and used them to update the dataset when we found some mistakes. Another threat is that Solidity is a fast-growing programming language. There are nine versions released in 2018, which may add or modify any features of the previous version. \textit{DefectChecker} is designed based on Solidity version 0.4.0+, which is the most popular version in the time of writing the paper~\cite{EtherScan}. In the future, more smart contracts may use higher versions, which may make our tool unable to work.

	\section{A Large Scale Evaluation} 

In the previous section, we showed that \textit{DefectChecker} has an excellent performance when applied to a small scale dataset. In this section, to validate \textit{DefectChecker} is still usable to find contract defects in real-world smart contracts, we ran \textit{DefectChecker} on a large scale dataset that we crawled from Ethereum blockchain, and show the contract defects as found by  \textit{DefectChecker}. We give two real-world attacks as case studies to show how harmful these contract defects are.

\subsection{Dataset} 
To identify whether contract defects are actually prevalent in a large-scale, real-world dataset, we crawled bytecode from Ethereum blockchain by 2019.01 and obtained 183,706 distinct bytecode. Since some smart contract versions are not supported by \textit{DefectChecker}, and so we removed them from our experimental dataset. Finally, we ran \textit{DefectChecker} on 165,621 distinct smart contract bytecode. All these bytecode are runtime bytecode. Runtime bytecode does not contain information on their constructor function. It is the default bytecode stored on the Ethereum.

\subsection{Contract Defects on Ethereum} 

We ran \textit{DefectChecker} on 165,621 smart contract bytecode. The detailed results are given in Table~\ref{tab:percent}, which aims to show the frequency of each defect on Ethereum. Since \textit{DefectChecker}  only identifies whether a contract contains a defect or not,  if the same kind of defects appears multiple times in a smart contract, we only count it once in Table~\ref{tab:percent}. The second column of the table shows how many contracts contain related defects, and the last column gives the percentage of how many contracts contain the defect. If a contract contains multiple defects, all of the defects are counted. 

\textit{Unchecked External Calls} is the most frequent contract defect in the Ethereum, and about 7.5\% of real world smart contracts contain this defect. There are about 3.1\% of smart contracts that contain \textit{Block Info Dependency}, which is the second most popular contract defect on the blockchain. \textit{Strict Balance Equality} is the rarest of our contract defects. \textit{DefectChecker} only detects 390 smart contracts that have this contract defect. The percentage of \textit{Nested Call} is also less than 1\%, with 1,043 (0.6\%) smart contracts having this kind of contract defect. The percentage of \textit{Transaction State Dependency}  and \textit{DoS Under External Influence}  are similar on Ethereum, at about 1.0\% and 1.3\%, respectively. There are 3,139 greedy contracts on the Ethereum, and 3,892 smart contracts containing the \textit{Reentrancy} problem, which can lead to serious security problems. 

\begin{table}
	\small
	
	\caption{Contract Defects in Ethereum} 
	\label{tab:percent}
	\centering
	\begin{tabular}{l | r | r  }
		\hline
		\textbf{Contract Defects} & \textbf{\# Defects} & \textbf{\# Percentage}  \\
		\hline
		\textit{Transaction State Dependency}  & 1,669 & 1.0\% \\
		\hline
		\textit{DoS Under External Influence}  & 2,116 & 1.3\%  \\
		\hline
		\textit{Strict Balance Equality}   & 390 & 0.2\%   \\
		\hline
		\textit{Reentrancy} & 3,892 & 2.4\% \\
		\hline
		\textit{Nested Call}  & 1,043 & 0.6\% \\
		\hline
		\textit{Greedy Contract} & 3,139 & 1.9\% \\
		\hline
		\textit{Unchecked External Calls} & 12,439 & 7.5\% \\
		\hline
		\textit{Block Info Dependency}  & 5,201 & 3.1\% \\

	\end{tabular} 
	
\end{table}

We found that there are 16 smart contracts that contain 4 kinds of contract defects, which are thus the most defective contracts. The number of smart contracts that contain 3 kinds of contract defects is 539, and 3,520 smart contracts contain 2 kinds of contract defects. About 25,815 smart contracts contain at least one kind of defect, which means that about 15.9\% smart contracts on Ethereum contain some kinds of defects, as reported by our \textit{DefectChecker}.

\begin{figure}
	\begin{center}
		
		\includegraphics[width=0.48\textwidth]{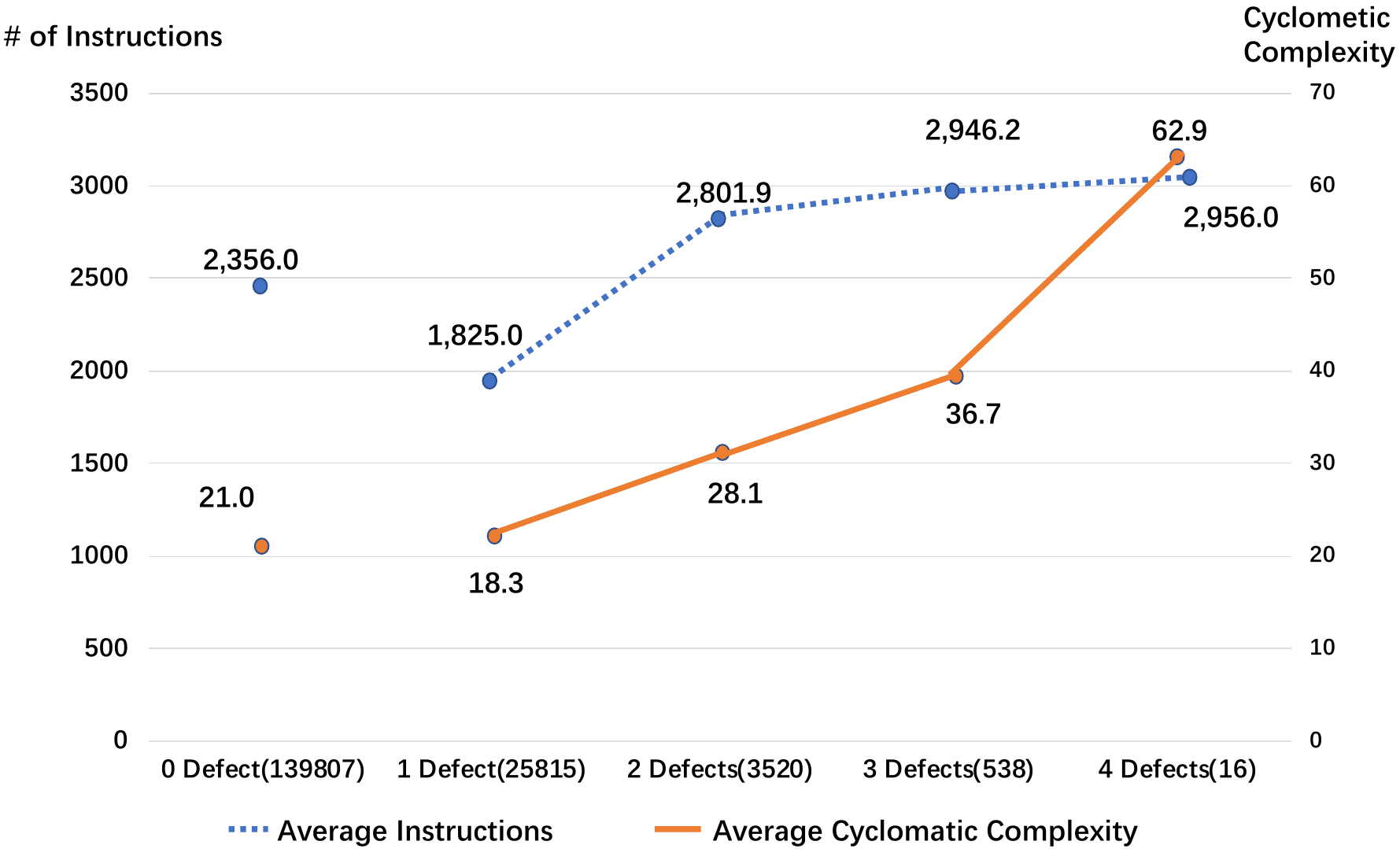} 
		
		\caption {The relationship between the number of contract defects and number of Instructions \& Cyclomatic Complexity}\vspace{-0.3cm}
		\label{Fig:cc}
	\end{center}   
\end{figure}

We utilized cyclomatic complexity~\cite{complexity} and the number of instructions to conduct a further analysis. We computed the cyclomatic complexity and number of instructions for contracts in our dataset. We found that the average cyclomatic complexity of smart contracts in Ethereum is 21.3, and the average number of instructions are 2,342.6.  Figure~\ref{Fig:cc} shows the relationship between the number of the contract defects that contained in smart contracts and the number of instructions \& cyclomatic complexity. The x-axis means the number of contract defects in a smart contract. The left y-axis is the number of x, and the right y-axis is the number of cyclomatic complexity. The two lines have a similar trend. 

The number of instructions is proportional to the length of a contracts' code, which can show the contracts' complexity at the code level. The number of cyclomatic complexity indicated the complexity of a program. We performed a generalized linear regression with the Poisson error distribution model provided by R~\cite{GLM} to analyze the relationship between the number of defects with instructions, and the number of defects with cyclomatic complexity. In our model, we use  the number of instructions and cyclomatic complexity to predict the number of defects, respectively. Since both the correlation coefficients are positive (0.001 with std. error = 0.0009 and 0.023 with std. error = 0.0179, respectively), it shows that the more complex a contract is, the higher is its probability to contain defects. We calculated the correlation level between these two complexity measures using the Pearson correlation method~\cite{benesty2009pearson} at a 5\% significance level. The statistical test shows that the correlation coefficient is 0.702 with $p-value < 0.05$. These correlation results imply that the number of instructions and cyclomatic complexity is correlated, and we can use only one of them as a predictor.

\subsection{Case Study}
\textit{DefectChecker} found some real-world attacks / financial loss from our large-scale testing on the full Ethereum dataset. In this subsection, we give two examples to show the importance of detecting such contract defects. 

\noindent \textbf{Case Study 1:} The first example is shown in Listing 14. There are 2,335.8 Ethers in the contract balance, and it is worth \$552,720 by Mar. 2020. Unfortunately, all the Ethers are locked because of the contract defect, i.e., \textit{Nested Call}.  The buggy function in Listing 14 is named \textit{sendReward()}. We highlight two lines of the code (Line 2 and Line 14), which are related to two contract defects, i.e., \textit{Nested Call} and \textit{DoS Under External Influence}. 

\begin{figure}
	\begin{center}
		
		\includegraphics[width=0.5\textwidth]{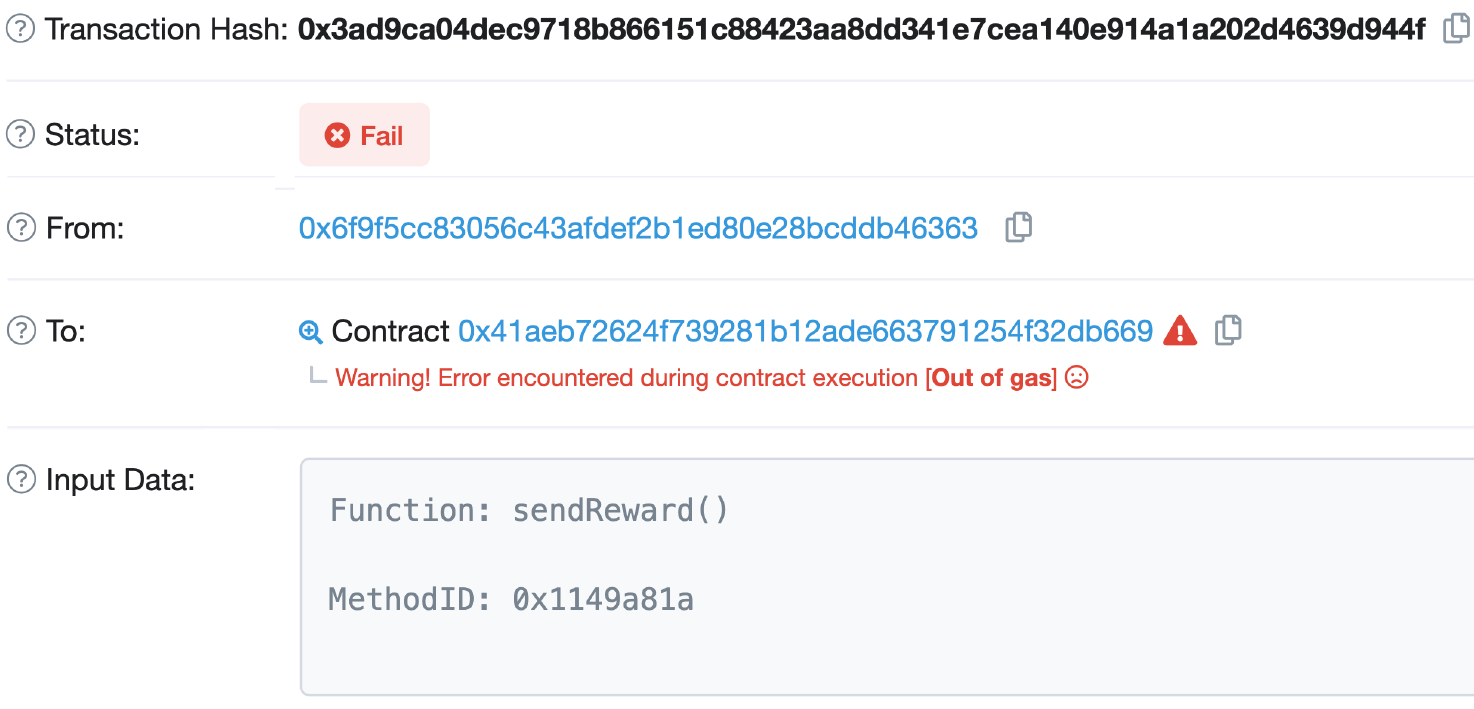} 
		
		\caption {Transaction Detail of Case Study 1}  \vspace{-0.1cm}
		\label{Fig:example1-trandetail}
	\end{center}   
\end{figure}

There is a loop in the function \textit{sendReward()}, and the loop iterations are increased with the length of \textit{investors[]}. However, the contract does not limit its loop iterations. As we know, sending Ethers is expensive as it needs a large amount of gas consumption, and the contract sends Ethers to the contract users in Line 14. So, the gas consumption of executing \textit{sendReward()} will increase in the length of \textit{investors[]}. When we check the transaction of the contract, we can find that the contract can work normally at first, as the total gas consumption of \textit{sendReward()} does not exceed its maximum gas limitation at that time. However, with the increase of the length of \textit{investors[]}, the total gas cost increases rapidly. The gas cost then eventually exceeds the gas limitation, and leads to an out of gas error. Even worse, since the length of \textit{investors[]} cannot be reduced, once the error happens, the \textit{sendReward()} cannot be called anymore, which means all the Ethers in the balance are locked forever. Figure~\ref{Fig:example1-trandetail} shows the detail of a failed transaction. It is clear that when a user calls \textit{sendReward()}, the out-of-gas error happens. 

\begin{lstlisting}[caption={Case Study 1 - Contract with Nested Call. Code from Contract: 0x41AeB72624f739281b12aDE663791254F32DB669. }]
function sendReward() public isOwner{
  for (uint i = 0; i < investors.length; i ++){
	address _add = investors[i];
	User memory _user = addressToUser[_add];
	if (_user.gameOver){
		autoReInvest(_add);
		_user.rebirth = now - (oneLoop / 2);
		addressToUser[_add] = _user;
	}else {
		if (SafeMath.sub(now , _user.rebirth) >=  oneLoop){
 			address payable needPay = address(uint160(_add));
			uint staticAmount = getStatic(_add);
			if (staticAmount > 0){
				needPay.transfer(staticAmount);
			}
	...
	}
}
\end{lstlisting}

It should be noticed that although the financial loss in the real world example is caused by \textit{Nested Call}, the contract shown in Listing 14 also has another contract defect, namely \textit{DoS Under External Influence}. This contract defect can also lead to the lock of Ethers. Specifically, if the \textit{needPay} (Line 14) is a contract address, the maximum \textit{Gas Limit} will be restricted to 2300 gas units, which is not enough to transfer Ethers. Thus, an \textit{out-of-gas error} will happen in Line 14, and the Ether transfer cannot succeed. 

%
%

\noindent \textbf{Case Study 2:} A second example is a bank contract, which is shown in Listing 15. Users can send Ethers to the \textit{Deposit()} function, and withdraw its Ethers by calling the \textit{CashOut()} function. First, the contract sends Ethers on Line 11 and then reduce the caller's balance on Line 12. However, it can lead to the \textit{Reentrancy} if the caller is an attacking contract. When the victim contract sends Ethers to the attack contract. The fallback function of the attack contract can recall the \textit{CashOut()} function, and steal Ethers of the victim contract. Then, all of the balance in the contract was stolen by the attackers. 

\begin{figure}
	\begin{center}
		
		\includegraphics[width=0.5\textwidth]{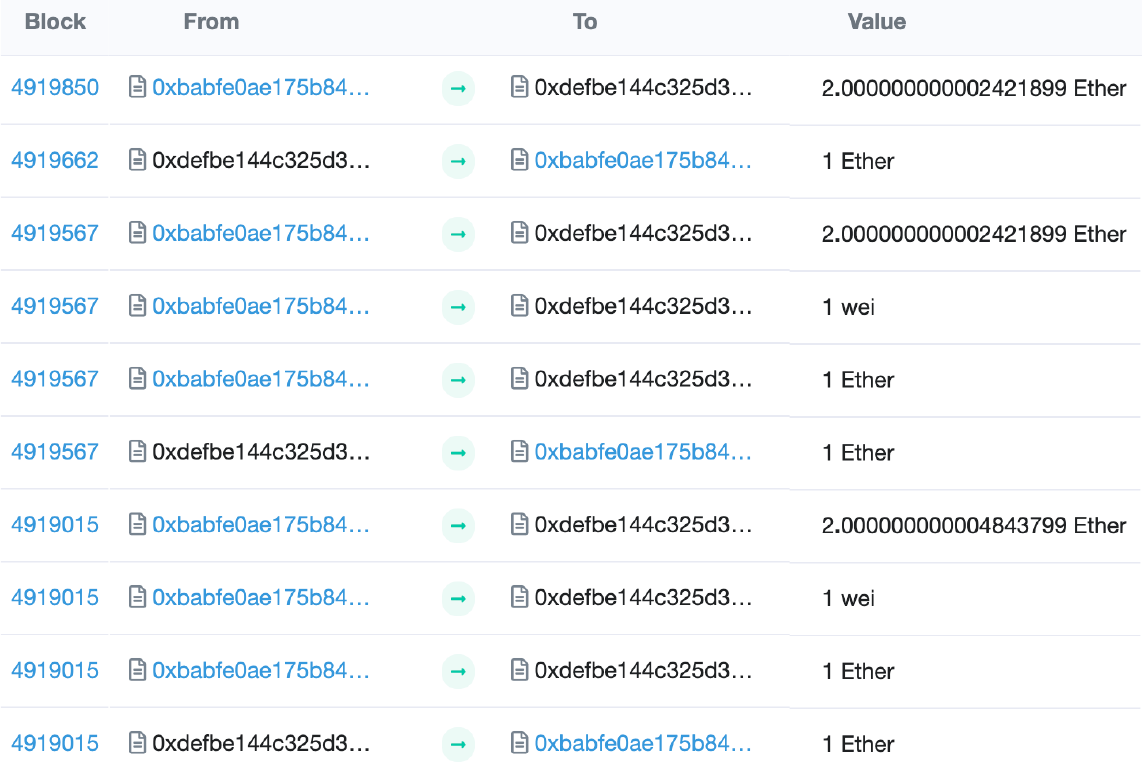} 
		
		\caption {Transaction Lists of Case Study 2} \vspace{-0.1cm}
		\label{Fig:example2-trans}
	\end{center}   
\end{figure}

Figure~\ref{Fig:example2-trans} shows an attacking transaction which was launched by an attacking contract. The address of the attacking contract starts with 0xdefbe, and the address of the victim contract starts with 0xbabfe. The attack happens three times on block 4919015, 4919567, and 4919662, respectively. First, the attacking contract sent 1 Ether to the victim contract. Then, the victim contract returned back Ethers to the attack contract. From these 3 attacks, the attacking contract stole about 5 Ethers from the victim contract, which were worth about \$1,200 at the time of writing the paper. We only show one example in Figure~\ref{Fig:example2-trans}. Actually, the victim contract was attacked by multiple attacking contracts, so the financial loss was far more than 5 Ethers. 

%
%

\begin{lstlisting}[caption={Case Study 2 - Contract with Reentrancy. Code from Contract: 0xbABfE0AE175b847543724c386700065137d30e3B. }]
function Deposit() public payable{
  if(msg.value >= MinDeposit){
	balances[msg.sender]+=msg.value;
	TransferLog.AddMessage(msg.sender,msg.value,"Deposit");
  }
}

function CashOut(uint _am)
{
  if(_am<=balances[msg.sender]){
	if(msg.sender.call.value(_am)()){
		balances[msg.sender]-=_am;
		TransferLog.AddMessage(msg.sender,_am,"CashOut");
	}
  }
}
\end{lstlisting}

\subsection{Threats to Validity} 

\textbf{Internal Validity.}
The dataset we used was crawled from Ethereum, which contains different Solidity versions. \textit{DefectChecker} only supports versions higher than 0.4.0+, and about 20,000 contracts had to be removed from our dataset, which may influence the overall results. However, the bytecode we removed is from many years ago, since the first version of 0.4.0+ was released on Sept. 2016. Even though there are many contract defects in the removed bytecode, these do not represent current smart contract usage. 

Another key threat is that we used our \textit{DefectChecker} to get the results, but \textit{DefectChecker} also reports false positives and negatives, as shown in the previous section. However, \textit{DefectChecker} is the most accurate and efficient tool that detects contract defects in the bytecode level, as we also demonstrated in the previous section. Therefore, we believe the results and our conclusions from it are reasonable. 

\textbf{External Validity.} 
There are more than 1,000 smart contracts being deployed to Ethereum every day~\cite{chen2019dataether}. Many guidance and security detection tools~\cite{oyente-tools, Maian} are released to the public, which can help to improve the quality of smart contracts. In this case, the contract defects in smart contracts may decrease, which may lead to different results to what we found and reported in this section.

	\section{Related Work}
\label{related}

\noindent {\bf Contract Defects on Smart Contracts. }
Our previous work~\cite{fse-smell} is the first work that defines 20 smart contract defects on Ethereum by analyzing the post on StackExchange~\cite{StackExchange}. We first crawl all 17,128 Stack Exchange posts by the time of writing the paper and use key words to filter solidity related posts. After getting Solidity related posts, two authors of the paper use \textit{Open Card Sorting} to find 20 contract defects and divide them into five categories, i.e., \textit{security, availability, performance, maintainability}, and \textit{reusability defects}. According to their paper, although previous works define several security defects, they did not consider the practitioners' perspective. Therefore, we first designed an online survey to collect feedback from developers to validate whether the developers regard the contract defects are harmful. This feedback showed that all the defined contract defects are harmful to smart contracts. We assigned five impact levels to the defined 20 contract defects according to our survey results and the symptoms of the defects. According to our definition, contract defects with impact level 1-3 can lead to unwanted behaviors of contract, e.g., a contract being controlled by attackers. 


\noindent {\bf Smart Contract Security Problems and Detection Tools. }
Luu et al. ~\cite{oyente} introduced four security issues in their work, i.e., mishandled exception, transaction-ordering dependence, timestamp dependence, and reentrancy attack. They proposed a tool named \textit{Oyente}, which is the first symbolic execution based bug detection tool for smart contracts.  They first split the bytecode into several blocks, and built a skeletal control flow graph for the detected contract. Then, they utilized Z3~\cite{z3} as their SMT solver and symbolically executed each instruction to obtain the full control flow graph. Finally, they designed different patterns to detect whether the input contracts contain the defined security problems. Oyente measured 19,366 existing Ethereum contracts and found 8,519 of them contain the defined security problems.

Kalra et al.~\cite{Zeus} developed a tool named Zeus. The tool feeds source code as input and translates them to LLVM bytecode. Zeus can detect seven kinds of security problems (four of them are the same with Oyente), and the other three problems are \emph{unchecked send, Failed send, Integer overflow/underflow}. They also compared their result to Oyente and found Oyente contains many false positives and false negatives. Zeus crawled 1,524 distinct smart contracts from Etherscan~\cite{EtherScan}, Etherchain~\cite{EtherChain} and EtherCamp~\cite{EtherCamp} explorers to evaluate their tool. The result illustrates that about 94.6\% of contracts contain at least one security problem. However, the needs of source code limited their usage.

Jiang et al.~\cite{Contractfuzzer} proposed a tool named \textit{ContractFuzzer} to test seven security issues.  \textit{ContractFuzzer} is the first tool that utilizes fuzzing technology to detect security problems on smart contracts. They tested 6,991 smart contracts and found that 459 of them have issues. However, only less than 0.5\% of smart contracts open their ABI to investigate on Ethereum~\cite{EtherScan}, while their tool needs smart contract ABI or source code to generate test case, which limited their usage. In addition, our dataset consisted of 579 bytecode smart contracts, which are not supported by \textit{ContractFuzzer}.

Nikolic et al.~\cite{Maian} developed a tool named MAIAN, which contains two major parts: symbolic analysis and concrete validation. Similar to Oyente, MAIAN utilizes symbolic execution and defines several execution rules to detect these security issues. Their tool takes input data as either bytecode or source code.  MAIAN has a different concern compared to our tool. They focus on security issues that can lead to a contract not able to release Ethers, can transfer Ethers to arbitrary addresses, or can be killed by anybody. Their results were deduced from 970,898 smart contracts and they found that a total of 34,200 (2,365 distinct) contracts contain at least one of these three security issues.

ConsenSys is a leading blockchain technology company. They built a website named SWC Registry~\cite{SWC} (Smart Contract Weakness Classification and Test Cases) to collect smart contract security problems from both online posts and news through crowdsourcing. \textit{Mythril}~\cite{Mythril} is a tool to detect security problems on this SWC Registy, and their first version was released in May 2018. The method used by \textit{Mythril} is similar to \textit{Oyente}. It first builds a CFG and utilizes Z3~\cite{z3} as an SMT solver. Then, it designs several rules to detect related problems. \textit{Mythril} is a tool developed by industry; their instruction manual does not contain any evaluation section on the tool. 

\textit{Securify}~\cite{Securify} is a tool released by Tsankov et al. \textit{Securify} is the first tool that utilizes semantic information to detect security problems on smart contracts. It first decompiles EVM bytecode to and analyzes the semantic facts, including data flow and control flow dependencies. Finally, it checks several security patterns that are written in a specialized domain-specific language to detect related security problems. \textit{Securify} focuses on two kinds of security problems, i.e., \textit{Stealing Ether} and \textit{Frozen Funds}. There are 9 security issues can that be detected by \textit{Securify}. Tsankov et al. evaluate their tool based on two datasets. First, a large-scale evaluation based on 24,594 smart contracts. Their results show that more than 70\% of smart contracts contain at least one of the security problems. Then, they use a small-scale evaluation based on 100 smart contracts to evaluate their proposed tool's effectiveness. To simplify manual inspection, all of these 100 smart contracts are up to 200 lines of code. According to their paper, \textit{Securify} can find more security violations compared to \textit{Oyente} and \textit{Mythril}.

In this paper, we propose a tool named \textit{DefectChecker}, which is the most accurate and the fastest symbolic execution model of smart contract defect detection tool. \textit{DefectChecker} can detect contract defects by analyzing bytecode, while  \textit{Zeus} and \textit{ContractFuzzer} need source code and contract ABI, respectively. The bytecode of smart contracts are visible to everyone, while only 1\% of smart contracts open up their source code and ABI for the public~\cite{TokenScope}, which restricts their usage. \textit{MAIAN} uses a dynamic analysis method to detect security problems, which is different from our static analysis method. However, we find their tool can not support the current version of Ethereum that we used. \textit{Oyente, Mythril}, and \textit{Securify} use symbolic execution to detect security problems, which are similar to \textit{DefectChecker}, but \textit{DefectChecker} uses \textit{Stack Event} and \textit{Feature Detector} to instead the usage of SMT solver, which makes \textit{DefectChecker} requires less runtime and yet is more accurate than these tools. 

\textit{Oyente, Mythril,} and \textit{Securify} can detect other contract defects that are not supported by \textit{DefectChecker}. Especially for \textit{Mythril}, which can detect 34 kinds of contract defects. We admit that some tools can detect more contract defects than \textit{DefectChecker}, but it is not the main motivation of this paper.  Previous works, e.g., \textit{Oyente, Securify},  only proposed several security defects of smart contracts without validating they are really harmful. This is not beneficial for the development of the smart contract ecosystem. In our previous work, we validated whether smart contract developers consider the contract defects we found from StackExchange posts are harmful by using an online survey. In this paper, we proposed \textit{DefectChecker}, which aims to automatically detect the validated contract defects. We use \textit{Oyente, Mythril,} and \textit{Securify} as baseline methods with the aim to show the method we use is more accurate and efficient than these state-of-the-art tools.

 Our \textit{DefectChecker} is extensible. As shown in Figure~\ref{Fig:overall}, there are three components of \textit{DefectChecker}, i.e., \textit{CFG Builder}, \textit{Feature Detector}, and \textit{Defect Identifier}. \textit{Defect Identifier} uses eight different rules to identify the contract defects, while the other two components can also be used to detect other defects. When detecting other defects, we can define new rules that use the data provided by our Feature Detector, CFG, and Stack Event components. There are many tools built based on the top of Oyente. For example, our previous work \textit{GasChecker}~\cite{chen2020gaschecker} is a tool to detect gas-inefficient Smart Contracts. The tool uses the CFG generated by \textit{Oyente} to  detect related gas-inefficient issues. \textit{DefectChecker} has higher efficiency in generating CFG compared to \textit{Oyente}. \textit{GasChecker} can also use the CFG generated by \textit{DefectChecker}.  Thus, \textit{DefectChecker} is also extensible to detect other kinds of issues.

	\section{Conclusion and Future Work}

In this paper, we proposed \textit{DefectChecker}, which utilizes symbolic execution to detect smart contract defects by analyzing the contracts' bytecode. \textit{DefectChecker} uses different rules to detect 8 contract defects and achieves a very good result when running on our previous work's dataset. The scores for our tool are much higher than those of the state of the art work e.g. (\textit{Oyente, Mythril}, and \textit{Securify}). We also crawled 165,621 distinct bytecode smart contracts from Ethereum and ran \textit{DefectChecker} on these. Our results show that about 15.89\% of smart contracts on Ethereum contain at least one instance of our 8 identified kinds of contract defects. 

Two groups can benefit from this work. For smart contract developers, they can utilize \textit{DefectChecker} to check their smart contracts and make them more robust. As \textit{DefectChecker} can detect contract defects from bytecode without the need for source code, developers can utilize \textit{DefectChecker} to check whether the smart contracts they call are secure or not, even if the callee contracts are not open sourced. This can also make their contracts safer. For software engineering researchers, \textit{DefectChecker} provides a good framework to help them solve other smart-contract-related research problems as the CFG generated by \textit{DefectChecker} can be used for other purposes.

\textit{DefectChecker} has some false positives / negatives when detecting defects, e.g., \textit{NC, DuEI}. ﻿ As we described in Section 4.4, adding a SMT Solver can reduce some error cases, while it will also increase the time consumption for analyzing a contract. Future work could explore how to combine the method used by \textit{DefectChecker} and a SMT solver, to balance both efficiency and accuracy. Specifically, researchers could identify which kinds of code patterns can lead to the errors made by \textit{DefectChecker}. For example, \textit{DefectChecker} regards all paths to be reachable, while some conditional expressions are always evaluated to false, which can lead to the false positives in detecting loops. Developers can use a SMT solver to check the conditional expression in the loop related blocks. This method can increase the accuracy in detecting loop related blocks.

\section*{Acknowledgements}

This research was partially supported by the Australian Research Council's Discovery Early Career Researcher Award (DECRA) funding scheme (DE200100021), ARC Laureate Fellowship funding scheme (FL190100035), ARC Discovery grant (DP200100020), National Natural Science Foundation of China (61872057),  National Key R\&D Program of China (2018YFB0804100), Hong Kong RGC Project (No. 152193/19E), and the National Research Foundation, Singapore under its Industry Alignment Fund – Pre-positioning (IAF-PP) Funding Initiative. Any opinions, findings and conclusions or recommendations expressed in this material are those of the author(s) and do not reflect the views of National Research Foundation, Singapore.

	\balance

	\bibliographystyle{IEEEtran}
	\bibliography{ref}

\begin{thebibliography}{10}
\providecommand{\url}[1]{#1}
\csname url@samestyle\endcsname
\providecommand{\newblock}{\relax}
\providecommand{\bibinfo}[2]{#2}
\providecommand{\BIBentrySTDinterwordspacing}{\spaceskip=0pt\relax}
\providecommand{\BIBentryALTinterwordstretchfactor}{4}
\providecommand{\BIBentryALTinterwordspacing}{\spaceskip=\fontdimen2\font plus
\BIBentryALTinterwordstretchfactor\fontdimen3\font minus
  \fontdimen4\font\relax}
\providecommand{\BIBforeignlanguage}[2]{{%
\expandafter\ifx\csname l@#1\endcsname\relax
\typeout{** WARNING: IEEEtran.bst: No hyphenation pattern has been}%
\typeout{** loaded for the language `#1'. Using the pattern for}%
\typeout{** the default language instead.}%
\else
\language=\csname l@#1\endcsname
\fi
#2}}
\providecommand{\BIBdecl}{\relax}
\BIBdecl

\bibitem{bitcoin}
S.~Nakamoto, ``Bitcoin: A peer-to-peer electronic cash system,'' \emph{Working
  Paper}, 2008.

\bibitem{Ethereum_yellow_paper}
G.~Wood, ``Ethereum: A secure decentralised generalised transaction ledger,''
  \emph{Project Yellow Paper}, 2014.

\bibitem{dapp}
\BIBentryALTinterwordspacing
(Apr., 2019) {Decentralized Application}. [Online]. Available:
  \url{https://en.wikipedia.org/wiki/Decentralized\_application}
\BIBentrySTDinterwordspacing

\bibitem{cryptokitties}
\BIBentryALTinterwordspacing
(Feb., 2019) Cryptokitties. [Online]. Available:
  \url{https://www.cryptokitties.co/}
\BIBentrySTDinterwordspacing

\bibitem{ICO}
\BIBentryALTinterwordspacing
(Apr., 2019) {ICO Ethereum}. [Online]. Available:
  \url{https://etherscan.io/directory/ICOs}
\BIBentrySTDinterwordspacing

\bibitem{Solidity}
\BIBentryALTinterwordspacing
(Mar., 2018) {Solidity Document}. [Online]. Available:
  \url{http://solidity.readthedocs.io}
\BIBentrySTDinterwordspacing

\bibitem{chen2017adaptive}
T.~Chen, X.~Li, Y.~Wang, J.~Chen, Z.~Li, X.~Luo, M.~H. Au, and X.~Zhang, ``{An
  Adaptive Gas Cost Mechanism for Ethereum to Defend Against Under-Priced DoS
  Attacks},'' in \emph{International Conference on Information Security
  Practice and Experience}.\hskip 1em plus 0.5em minus 0.4em\relax Springer,
  2017, pp. 3--24.

\bibitem{chen2018understanding}
T.~Chen, Y.~Zhu, Z.~Li, J.~Chen, X.~Li, X.~Luo, X.~Lin, and X.~Zhange,
  ``{Understanding Ethereum via Graph Analysis},'' in \emph{IEEE INFOCOM
  2018-IEEE Conference on Computer Communications}.\hskip 1em plus 0.5em minus
  0.4em\relax IEEE, 2018, pp. 1484--1492.

\bibitem{chen2020understanding}
T.~Chen, Z.~Li, Y.~Zhu, J.~Chen, X.~Luo, J.~C.-S. Lui, X.~Lin, and X.~Zhang,
  ``{Understanding Ethereum via Graph Analysis},'' \emph{ACM Transactions on
  Internet Technology (TOIT)}, vol.~20, no.~2, pp. 1--32, 2020.

\bibitem{iso2017iec}
{ISO}, ``{ISO/IEC/IEEE International Standard - Systems and software
  engineering--Vocabulary},'' ISO/IEC/IEEE 24765: 2017 (E), Tech. Rep., 2017.

\bibitem{fse-smell}
J.~Chen, X.~Xia, D.~Lo, J.~Grundy, X.~Luo, and T.~Chen, ``{Defining Smart
  Contract Defects on Ethereum},'' \emph{IEEE Transactions on Software
  Engineering}, 2020.

\bibitem{chillarege1996orthogonal}
R.~Chillarege \emph{et~al.}, ``Orthogonal defect classification,''
  \emph{Handbook of Software Reliability Engineering}, pp. 359--399, 1996.

\bibitem{StackExchange}
\BIBentryALTinterwordspacing
(Jan., 2018) {StackExchange}. [Online]. Available:
  \url{https://ethereum.stackexchange.com/}
\BIBentrySTDinterwordspacing

\bibitem{chen2018towards}
T.~Chen, Z.~Li, H.~Zhou, J.~Chen, X.~Luo, X.~Li, and X.~Zhang, ``{Towards
  saving money in using smart contracts},'' in \emph{2018 IEEE/ACM 40th
  International Conference on Software Engineering: New Ideas and Emerging
  Technologies Results (ICSE-NIER)}.\hskip 1em plus 0.5em minus 0.4em\relax
  IEEE, 2018, pp. 81--84.

\bibitem{whitepaper}
{Ethereum Foundation}, ``Ethereum’s white paper.'' \emph{https:
  //github.com/ethereum/wiki/wiki/White-Paper}, 2014.

\bibitem{soot}
R.~Vall{\'e}e-Rai, P.~Co, E.~Gagnon, L.~Hendren, P.~Lam, and V.~Sundaresan,
  ``{Soot: A Java bytecode optimization framework},'' in \emph{CASCON First
  Decade High Impact Papers}, 2010, pp. 214--224.

\bibitem{grech2019gigahorse}
N.~Grech, L.~Brent, B.~Scholz, and Y.~Smaragdakis, ``{Gigahorse: thorough,
  declarative decompilation of smart contracts},'' in \emph{{2019 IEEE/ACM 41st
  International Conference on Software Engineering (ICSE)}}.\hskip 1em plus
  0.5em minus 0.4em\relax IEEE, 2019, pp. 1176--1186.

\bibitem{jvm}
\BIBentryALTinterwordspacing
(Aug., 2020) {The Java Virtual Machine Specification}. [Online]. Available:
  \url{https://docs.oracle.com/javase/specs/jvms/se8/html/index.html}
\BIBentrySTDinterwordspacing

\bibitem{chen2019large}
T.~Chen, Z.~Li, Y.~Zhang, X.~Luo, T.~Wang, T.~Hu, X.~Xiao, D.~Wang, J.~Huang,
  and X.~Zhang, ``{A large-scale empirical study on control flow identification
  of smart contracts},'' in \emph{{2019 ACM/IEEE International Symposium on
  Empirical Software Engineering and Measurement (ESEM)}}.\hskip 1em plus 0.5em
  minus 0.4em\relax IEEE, 2019, pp. 1--11.

\bibitem{oyente}
L.~Luu, D.-H. Chu, H.~Olickel, P.~Saxena, and A.~Hobor, ``Making smart
  contracts smarter,'' in \emph{Proceedings of the 2016 ACM SIGSAC Conference
  on Computer and Communications Security}.\hskip 1em plus 0.5em minus
  0.4em\relax ACM, 2016, pp. 254--269.

\bibitem{geth}
\BIBentryALTinterwordspacing
(Mar., 2018) Geth. [Online]. Available:
  \url{https://github.com/ethereum/go-ethereum}
\BIBentrySTDinterwordspacing

\bibitem{TokenScope}
T.~Chen, Y.~Zhang, Z.~Li, X.~Luo, T.~Wang, R.~Cao, X.~Xiao, and X.~Zhang,
  ``{TokenScope: Automatically Detecting Inconsistent Behaviors of currency
  Tokens in Ethereum},'' in \emph{Proceedings of the 2019 ACM SIGSAC Conference
  on Computer and Communications Security}, 2019, pp. 1503--1520.

\bibitem{chen2017under}
T.~Chen, Y.~Feng, Z.~Li, H.~Zhou, X.~Luo, X.~Li, X.~Xiao, J.~Chen, and
  X.~Zhang, ``Gaschecker: Scalable analysis for discovering gas-inefficient
  smart contracts,'' \emph{IEEE Transactions on Emerging Topics in Computing},
  2020.

\bibitem{SMT}
C.~Barrett and C.~Tinelli, ``Satisfiability modulo theories,'' in
  \emph{Handbook of Model Checking}.\hskip 1em plus 0.5em minus 0.4em\relax
  Springer, 2018, pp. 305--343.

\bibitem{z3}
L.~De~Moura and N.~Bj{\o}rner, ``{Z3: An efficient SMT solver},'' in
  \emph{International conference on Tools and Algorithms for the Construction
  and Analysis of Systems}.\hskip 1em plus 0.5em minus 0.4em\relax Springer,
  2008, pp. 337--340.

\bibitem{DFS}
R.~Tarjan, ``Depth-first search and linear graph algorithms,'' \emph{SIAM
  journal on computing}, vol.~1, no.~2, pp. 146--160, 1972.

\bibitem{EIP55}
\BIBentryALTinterwordspacing
(Jan., 2016) {EIP-55}. [Online]. Available:
  \url{https://github.com/ethereum/EIPs/blob/master/EIPS/eip-55.md}
\BIBentrySTDinterwordspacing

\bibitem{complexity}
T.~J. McCabe, ``A complexity measure,'' \emph{IEEE Transactions on software
  Engineering}, no.~4, pp. 308--320, 1976.

\bibitem{kappa}
J.~Cohen, ``A coefficient of agreement for nominal scales,'' \emph{Educational
  and psychological measurement}, vol.~20, no.~1, pp. 37--46, 1960.

\bibitem{EtherScan}
\BIBentryALTinterwordspacing
(Mar., 2018) Etherscan. [Online]. Available: \url{https://etherscan.io/}
\BIBentrySTDinterwordspacing

\bibitem{Maian}
I.~Nikoli{\'c}, A.~Kolluri, I.~Sergey, P.~Saxena, and A.~Hobor, ``Finding the
  greedy, prodigal, and suicidal contracts at scale,'' in \emph{Proceedings of
  the 34th Annual Computer Security Applications Conference}.\hskip 1em plus
  0.5em minus 0.4em\relax ACM, 2018, pp. 653--663.

\bibitem{Securify}
P.~Tsankov, A.~Dan, D.~Drachsler-Cohen, A.~Gervais, F.~Buenzli, and M.~Vechev,
  ``Securify: Practical security analysis of smart contracts,'' in
  \emph{Proceedings of the 2018 ACM SIGSAC Conference on Computer and
  Communications Security}.\hskip 1em plus 0.5em minus 0.4em\relax ACM, 2018,
  pp. 67--82.

\bibitem{Mythril}
\BIBentryALTinterwordspacing
(Aug., 2019) {Mythril: Security analysis tool for EVM bytecode. }. [Online].
  Available: \url{https://github.com/ConsenSys/mythril}
\BIBentrySTDinterwordspacing

\bibitem{Contractfuzzer}
B.~Jiang, Y.~Liu, and W.~Chan, ``Contractfuzzer: Fuzzing smart contracts for
  vulnerability detection,'' in \emph{Proceedings of the 33rd ACM/IEEE
  International Conference on Automated Software Engineering}.\hskip 1em plus
  0.5em minus 0.4em\relax ACM, 2018, pp. 259--269.

\bibitem{Zeus}
S.~Kalra, S.~Goel, M.~Dhawan, and S.~Sharma, ``Zeus: Analyzing safety of smart
  contracts,'' in \emph{25th Annual Network and Distributed System Security
  Symposium (NDSS’18)}, 2018.

\bibitem{kitchenham2004procedures}
B.~Kitchenham, ``Procedures for performing systematic reviews,'' \emph{Keele,
  UK, Keele University}, vol.~33, no. 2004, pp. 1--26, 2004.

\bibitem{Web3py}
\BIBentryALTinterwordspacing
(April., 2019) Web3.py. [Online]. Available:
  \url{https://web3py.readthedocs.io/en/stable/}
\BIBentrySTDinterwordspacing

\bibitem{GLM}
\BIBentryALTinterwordspacing
(Jan., 2021) {Generalized Linear Models in R, Part 6: Poisson Regression for
  Count Variables}. [Online]. Available:
  \url{https://www.theanalysisfactor.com/generalized-linear-models-in-r-part-6-poisson-regression-count-variables/}
\BIBentrySTDinterwordspacing

\bibitem{benesty2009pearson}
J.~Benesty, J.~Chen, Y.~Huang, and I.~Cohen, ``Pearson correlation
  coefficient,'' in \emph{Noise reduction in speech processing}.\hskip 1em plus
  0.5em minus 0.4em\relax Springer, 2009, pp. 1--4.

\bibitem{chen2019dataether}
T.~Chen, Z.~Li, Y.~Zhang, X.~Luo, A.~Chen, K.~Yang, B.~Hu, T.~Zhu, S.~Deng,
  T.~Hu \emph{et~al.}, ``{Dataether: Data Exploration Framework for
  Ethereum},'' in \emph{2019 IEEE 39th International Conference on Distributed
  Computing Systems (ICDCS)}.\hskip 1em plus 0.5em minus 0.4em\relax IEEE,
  2019, pp. 1369--1380.

\bibitem{oyente-tools}
\BIBentryALTinterwordspacing
(Mar., 2018) {Oyente: An Analysis Tool for Smart Contracts}. [Online].
  Available: \url{https://github.com/melonproject/oyente}
\BIBentrySTDinterwordspacing

\bibitem{EtherChain}
\BIBentryALTinterwordspacing
(Mar., 2018) Etherchain. [Online]. Available:
  \url{https://www.etherchain.org/contracts/}
\BIBentrySTDinterwordspacing

\bibitem{EtherCamp}
\BIBentryALTinterwordspacing
(Mar., 2018) Ethercamp. [Online]. Available: \url{https://live.ether.camp/}
\BIBentrySTDinterwordspacing

\bibitem{SWC}
\BIBentryALTinterwordspacing
(July., 2019) {SWC Registry: Smart Contract Weakness Classification and Test
  Cases}. [Online]. Available:
  \url{https://smartcontractsecurity.github.io/SWC-registry/}
\BIBentrySTDinterwordspacing

\bibitem{chen2020gaschecker}
T.~Chen, Y.~Feng, Z.~Li, H.~Zhou, X.~Luo, X.~Li, X.~Xiao, J.~Chen, and
  X.~Zhang, ``{GasChecker: Scalable Analysis for Discovering Gas-Inefficient
  Smart Contracts},'' \emph{IEEE Transactions on Emerging Topics in Computing},
  2020.

\end{thebibliography}
	
\end{document}